\documentclass[12pt,a4paper]{article} 

\usepackage{graphicx} 
\graphicspath{{H:/Pictures/}}
\newcommand{\nc}{\newcommand}
\nc{\etal}{{\it{et al. }}}
\usepackage{booktabs}

\nc{\ntitle}[1]{
 \begin{center}
   \fbox{\textbf{\Large #1}}
  \end{center}         }

\nc{\tred}[1]{\textcolor[rgb]{1.00,0.00,0.00}{#1}}
\nc{\tblue}[1]{\textcolor[rgb]{0.00,0.00,1.00}{#1}}

\nc{\slideline}{\smallskip \hrule\hrule \smallskip}
\nc{\stitle}[1]{
\textbf{\large #1}
\slideline
\slideline
          }

\nc{\nn}{\nonumber}
\nc{\fns}{\footnotesize}

\nc{\revisionline}{\vspace{.1in} \today \vspace{.1in} \hrule\hrule\hrule\vspace{.1in}}
\nc{\newpp}{\vspace{.1in} \noindent}

\nc{\wh}{\widehat}

\nc{\Ef}{ {\rm E}_{\infty} }
\nc{\Ex}{ {\rm E} }
\nc{\Ec}{ {\rm E}_1 }
\nc{\Pf}{ {\rm P}_{\infty} }
\nc{\Pc}{ {\rm P}_{1} }
\nc{\Prb}{ {\rm P} }
\nc{\sd}{\pm \hat{\sigma} }

\nc{\indep}{{\, \perp \! \! \! \perp  \,} }
\nc{\tsps}{^{ {\rm T} } }

\nc{\pu}{\pi_{\rm U}}
\nc{\pbi}{\pi_{\rm B}}
\nc{\pnb}{\pi_{\rm NB}}
\nc{\prp}{\propto}
\nc{\pr}{ {\rm pr} }

\nc{\al}{\alpha}
\nc{\dl}{\delta}
\nc{\la}{\lambda}
\nc{\om}{\omega}
\nc{\vep}{\varepsilon}

\nc{\snf}{\sum_{n=1}^{\infty}}
\nc{\skf}{\sum_{k=1}^{\infty}}
\nc{\sner}{\sum_{n=1}^{86}}
\nc{\sjn}{\sum_{j=1}^{n}}
\nc{\skn}{\sum_{k=1}^{n}}
\nc{\sumim}{\sum_{i=1}^m}
\nc{\sumjn}{\sum_{j=1}^n}
\nc{\sumlL}{\sum_{l=1}^{L}}
\nc{\sumL}{\sum_{l=1}^{L}}
\nc{\sumkK}{\sum_{k=1}^{K_i}}
\nc{\sumrR}{\sum_{r=1}^R}

\nc{\hivp}{\sum_{ {\rm HIV}^+ } }
\nc{\sumiN}{ \sum_{i=1}^N }
\nc{\summM}{ \sum_{m=1}^M }
\nc{\sumjM}{ \sum_{j=1}^M }
\nc{\lsq}{\left[}
\nc{\rsq}{\right]}
\nc{\lbc}{\left \{ }
\nc{\rbc}{\right \} }
\nc{\lp}{\left(}
\nc{\rp}{\right)}

\nc{\imp}{\Rightarrow}
\nc{\lbf}{\lim_{b \rightarrow \infty}}
\nc{\limNinf}{\lim_{N \rightarrow \infty}}
\nc{\limminf}{\lim_{m \rightarrow \infty}}
\nc{\limninf}{\lim_{n \rightarrow \infty}}
\nc{\convd}{\stackrel{D}{\longrightarrow}}
\nc{\convp}{\stackrel{P}{\longrightarrow}}
\nc{\eqd}{\stackrel{{\EuScript D}}{=}}

\nc{\trans}{^{\text T}}
\nc{\ol}{\overline}
\nc{\logit}{\text{logit}\,}

\nc{\rl}{ {\rm {\bf R} } }
\nc{\zah}{ {\rm {\bf Z} } }

\nc{\lkn}{\Lambda^n_k}
\nc{\stp}{ {\cal C}_b }
\nc{\istp}{ {\cal I}_A }
\nc{\snb}{S_{N_b}}
\nc{\stb}{S_{T_b}}
\nc{\ixlog}{I_{ \{ 0 \leq x \leq \log \al \} } }
\nc{\iulog}{I_{ \{ 0 \leq u  \leq \log \al \} } }
\nc{\rgn}{ \Upsilon_n }
\nc{\var}{{\rm var}}
\nc{\cov}{{\rm cov}}
\nc{\corr}{{\rm corr}}
\nc{\dpl}{\partial}
\nc{\half}{ {\textstyle \frac{1}{2}} }
\nc{\tr}{{\rm trace}}
\nc{\real}{\mathbb{R}}
\nc{\bbC}{\mathbb{C}}

\def\boxit#1{\vbox{\hrule\hbox{\vrule\kern6pt\vbox{\kern6pt#1\kern6pt}\kern6pt\vrule}\hrule}}

\nc{\calb}{ {\cal B} }
\nc{\calc}{ {\cal C} }
\nc{\bcalc}{ \mbox{\boldmath{${\cal C}$}}}
\nc{\cald}{ {\cal D} }
\nc{\cale}{ {\cal E} }
\nc{\cali}{ {\cal I} }
\nc{\call}{ {\cal L} }
\nc{\calm}{ {\cal M} }
\nc{\caln}{ {\cal N} }
\nc{\cals}{ {\cal S} }
\nc{\calo}{ {\cal O} }
\nc{\bcalo}{ \mbox{\boldmath{${\cal O}$}}}
\nc{\calt}{ {\cal T} }
\nc{\calv}{ {\cal V} }
\nc{\bcalu}{ \mbox{\boldmath{${\cal U}$}}}
\nc{\calu}{ {\cal U} }
\nc{\calw}{ {\cal W} }
\nc{\calx}{ {\cal X} }

\nc{\sca}{ {\EuScript A} }
\nc{\scb}{ {\EuScript B} }
\nc{\scc}{ {\EuScript C} }
\nc{\scd}{ {\EuScript D} }
\nc{\sce}{ {\EuScript E} }
\nc{\scf}{ {\EuScript F} }
\nc{\scF}{ {\EuScript f} }
\nc{\scg}{ {\EuScript G} }
\nc{\sch}{ {\EuScript H} }
\nc{\sci}{ {\EuScript I} }
\nc{\scj}{ {\EuScript J} }
\nc{\sck}{ {\EuScript K} }
\nc{\scl}{ {\EuScript L} }
\nc{\sclic}{ \scl_i^{\rm c} }
\nc{\scm}{ {\EuScript M} }
\nc{\scn}{ {\EuScript N} }
\nc{\sco}{ {\EuScript O} }
\nc{\scp}{ {\EuScript P} }
\nc{\scq}{ {\EuScript Q} }
\nc{\scr}{ {\EuScript R} }
\nc{\scs}{ {\EuScript S} }
\nc{\sct}{ {\EuScript T} }
\nc{\scu}{ {\EuScript U} }
\nc{\scv}{ {\EuScript V} }
\nc{\scw}{ {\EuScript W} }
\nc{\scx}{ {\EuScript X} }
\nc{\scy}{ {\EuScript Y} }
\nc{\scz}{ {\EuScript Z} }
\nc{\scxo}{ {\EuScript X}_{\rm obs} }
\nc{\Xobs}{ \pmb{\scx}_{\rm obs} }
\nc{\Xcom}{ \pmb{\scx} }
\nc{\Xmis}{ \pmb{\scx}_{\rm mis} }

\nc{\bsci}{ \mbox{\boldmath{$\sci$}}}
\nc{\bscj}{ \mbox{\boldmath{$\scj$}}}

\nc{\sumlic}{\sum_{l \in sclic}}

\nc{\scyo}{ {\EuScript Y}_{\rm obs} }

\nc{\bga}{\begin{array}{c}}
\nc{\ena}{\end{array}}

\nc{\mhat}{ {\hat{p}}_M }
\nc{\fhat}{ {\hat{p}}_F }
\nc{\ph} { \hat{p} }

\nc{\ta}{ {\tilde{a}} }
\nc{\tc}{ {\tilde{c}} }

\nc{\bal}{\mbox{\boldmath{$\alpha$}}}
\nc{\balpha}{\mbox{\boldmath{$\alpha$}}}
\nc{\bone}{\mbox{\boldmath{$1$}}}
\nc{\bbet}{\mbox{\boldmath{$\beta$}}}
\nc{\bbeta}{\mbox{\boldmath{$\beta$}}}
\nc{\bDel}{\mbox{\boldmath{$\Delta$}}}
\nc{\bDelta}{\mbox{\boldmath{$\Delta$}}}
\nc{\bdel}{\mbox{\boldmath{$\delta$}}}
\nc{\bdelta}{\mbox{\boldmath{$\delta$}}}
\nc{\bet}{\mbox{\boldmath{$\eta$}}}
\nc{\beps}{\mbox{\boldmath{$\epsilon$}}}
\nc{\bvep}{\mbox{\boldmath{$\vep$}}}
\nc{\bgam}{\mbox{\boldmath{$\gamma$}}}
\nc{\bgamma}{\mbox{\boldmath{$\gamma$}}}
\nc{\bGamma}{\mbox{\boldmath{$\Gamma$}}}
\nc{\bLam}{\mbox{\boldmath{$\Lambda$}}}
\nc{\bLambda}{\mbox{\boldmath{$\Lambda$}}}
\nc{\blambda}{\mbox{\boldmath{$\lambda$}}}
\nc{\bmu}{ \mbox{\boldmath{$\mu$}}}
\nc{\bOm}{ \mbox{\boldmath{$\Omega$}}}
\nc{\bOmega}{ \mbox{\boldmath{$\Omega$}}}
\nc{\bom}{ \mbox{\boldmath{$\omega$}}}
\nc{\bomega}{ \mbox{\boldmath{$\omega$}}}
\nc{\bpi}{ \mbox{\boldmath{$\pi$}}}
\nc{\bPi}{ \mbox{\boldmath{$\Pi$}}}
\nc{\bpsi}{ \mbox{\boldmath{$\psi$}}}
\nc{\bPsi}{ \mbox{\boldmath{$\Psi$}}}
\nc{\bphi}{ \mbox{\boldmath{$\phi$}}}
\nc{\bPhi}{ \mbox{\boldmath{$\Phi$}}}
\nc{\bxi}{ \mbox{\boldmath{$\xi$}}}
\nc{\bXi}{ \mbox{\boldmath{$\Xi$}}}
\nc{\bSig}{\mbox{\boldmath{$\Sigma$}}}
\nc{\bSigma}{\mbox{\boldmath{$\Sigma$}}}
\nc{\bsig}{\mbox{\boldmath{$\sigma$}}}
\nc{\btau}{\mbox{\boldmath{$\tau$}}}
\nc{\bThe}{\mbox{\boldmath{$\Theta$}}}
\nc{\bTheta}{\mbox{\boldmath{$\Theta$}}}
\nc{\bthe}{\mbox{\boldmath{$\theta$}}}
\nc{\btheta}{\mbox{\boldmath{$\theta$}}}
\nc{\bzeta}{\mbox{\boldmath{$\zeta$}}}
\nc{\bIm}{\mbox{\boldmath{$\Im$}}}

\nc{\ba}{ { \bf a }}
\nc{\bA}{ { \bf A }}
\nc{\bB}{ { \bf B }}
\nc{\bb}{ { \bf b }}
\nc{\bc}{ { \bf c }}
\nc{\bC}{ { \bf C }}
\nc{\bD}{ { \bf D }}
\nc{\bd}{ { \bf d }}
\nc{\be}{ { \bf e }}
\nc{\bF}{ { \bf F }}
\nc{\bG}{ { \bf G }}
\nc{\bh}{ { \bf h }}
\nc{\bH}{ { \bf H }}
\nc{\bI}{ { \bf I }}
\nc{\bJ}{ { \bf J }}
\nc{\bK}{ { \bf K }}
\nc{\bL}{ { \bf L }}
\nc{\bM}{ { \bf M }}
\nc{\bn}{ { \bf n }}
\nc{\bO}{ { \bf O }}
\nc{\bP}{ { \bf P }}
\nc{\br}{ { \bf r }}
\nc{\bR}{ { \bf R }}
\nc{\bs}{ { \bf s }}
\nc{\bS}{ { \bf S }}
\nc{\bT}{ { \bf T }}
\nc{\bt}{ { \bf t }}
\nc{\bu}{ { \bf u }}
\nc{\bU}{ { \bf U }}
\nc{\bv}{ { \bf v }}
\nc{\bV}{ { \bf V }}
\nc{\bW}{ { \bf W }}
\nc{\bw}{ { \bf w }}
\nc{\bx}{ { \bf x }}
\nc{\bX}{ { \bf X }}
\nc{\by}{ { \bf y }}
\nc{\bY}{ { \bf Y }}
\nc{\bz}{ { \bf z }}
\nc{\bZ}{ { \bf Z }}

\nc{\YR}{[\bY,R]}
\nc{\YgivenR}{[\bY \mid R]}
\nc{\RgivenY}{[R \mid \bY]}
\nc{\Y}{[\bY]}
\nc{\R}{[R]}

\nc{\dio}{d_i^o}
\nc{\timi}{t_{i,m_i}}
\nc{\betahat}{\hat{\bbet}}
\nc{\mui}{\bmu_{\rm I}}
\nc{\mue}{\bmu^{\rm E}}
\nc{\mup}{\bmu^{\rm P}}
\nc{\muihat}{\hat{\bmu}_{\rm I}}
\nc{\muehat}{\hat{\bmu}^{\rm E}}
\nc{\muphat}{\hat{\bmu}^{\rm P}}
\nc{\delhat}{\hat{\bdel}}
\nc{\muhat}{\hat{\bmu}}

\nc{\iid}{\stackrel{\rm iid}{\sim}}
\nc{\law}{\stackrel{\scl}{=}}

\nc{\phiij}{ \phi_{ij}( \Delta_0) }
\nc{\phiiprmj}{ \phi_{i'j}( \Delta_0) }
\nc{\phiijprm}{ \phi_{ij'}( \Delta_0) }
\nc{\phixy}{ \phi( X_i(S_{ik}), Y_j(T_{jl}) ) }
\nc{\phixydo}{ \phi( X_i(S_{ik}), Y_j(T_{jl})-\Delta_0 ) }
\nc{\phixyd}{ \phi( X_i(S_{ik}), Y_j(T_{jl})-\Delta) }
\nc{\phixydstar}{ \phi^*( X_i(S_{ik}), Y_j(T_{jl})-\Delta) }
\nc{\phixystdttil}{ \tilde{\phi}( X_i(s), Y_j(t)-\Delta, \theta) }
\nc{\phixydttil}{ \tilde{\phi}( X_i(S_{ik}), Y_j(T_{jl})-\Delta, \theta) }
\nc{\Nmn}{{\sqrt{N} \over mn}}

\nc{\Xis}{X_i(s)}
\nc{\Yjt}{Y_j(t)}

\nc{\bthehat}{\hat{\bthe}}

\nc{\Ritil}{\tilde{R}_i}

\nc{\Ybar}{\overline{Y}}
\nc{\Rbar}{\overline{R}}
\nc{\Nbar}{\overline{N}}
\nc{\intzeroinf}{\int_0^\infty}

\nc{\Fhat}{\hat{F}}
\nc{\Ghat}{\hat{G}}

\nc{\FhatS}{\hat{F}(S_{ik})}
\nc{\GhatT}{\hat{G}(T_{jl})}

\nc{\Fhatik}{\hat{F}_{ik}}
\nc{\Ghatjl}{\hat{G}_{jl}}
\nc{\Fik}{F_{ik}}
\nc{\Gjl}{G_{jl}}
\nc{\phiijkl}{\phi_{ik,jl}(\Delta)}
\nc{\phiijkltil}{\tilde{\phi}_{ik,jl}(\Delta_0,\theta_0)}
\nc{\ord}{N^{-3/2}}           % --- effective order of summation
\nc{\sumijkl}{\sum_{ijkl}}

\nc{\Citil}{\tilde{C}_i}
\nc{\Crtil}{\tilde{C}_r}
\nc{\Djtil}{\tilde{D}_j}
\nc{\Ditil}{\tilde{D}_i}

\nc{\Cithe}{\tilde{C}^{\theta}_i}
\nc{\Djthe}{\tilde{D}^{\theta}_j}

\nc{\Sikthe}{S_{ik}^{\theta}}
\nc{\Tjlthe}{T_{jl}^{\theta}}

\nc{\Zi}{ \bZ_{-i}}
\nc{\zic}{ \lbc z(\bs_j) \: : \: i \neq j \rbc }
\nc{\zkap}{ \bz_{\kappa} }
\nc{\sumi}{ \sum_i }
\nc{\sumj}{ \sum_j }
\nc{\sumij}{ \sum_{i < j} }
\nc{\sumiandj}{ \sum_{i, j} }
\nc{\zsi}{ z(\bs_i) }
\nc{\Zsi}{ Z(\bs_i) }
\nc{\zsj}{ z(\bs_j) }
\nc{\zsn}{ z(\bs_n) }
\nc{\zsone}{ z(\bs_1) }
\nc{\pZ}{ \Pr \lbc \bZ \rbc }
\nc{\qz}{ Q( \bz ) }
\nc{\qZ}{ Q( \bZ ) }

\nc{\thetaYD}{\theta_{Y\mid D}}
\nc{\thetaD}{\theta_D}
\nc{\psiDY}{\psi_{D\mid Y}}
\nc{\psiY}{\psi_Y}

\nc{\tn}{\Theta^{\nu}}
\nc{\Etn}{E_{\theta^{\nu}}}
\nc{\tnone}{\Theta^{\nu+1}}
\nc{\Lm}{L_{\text{m}}}
\nc{\Lo}{L_{\text{o}}}
\nc{\Ym}{Y_{\text{m}}}
\nc{\Yo}{Y_{\text{o}}}
\nc{\ym}{y_{\text{m}}}
\nc{\yo}{y_{\text{o}}}

\nc{\vijb}{v_{ij} - \bX_{i(j)}  \bbet}
\nc{\vikb}{v_{ik} - \bX_{i(k)}  \bbet}
\nc{\vilb}{v_{il} - \bX_{i(l)}  \bbet}
\nc{\betart}{ \bbet^{(r)}_{t_i} }
\nc{\betarj}{ \bbet^{(r)}_j }
\nc{\yij}{y_{ij}}
\nc{\Xmisi}{ {\bX_{ i{\rm (mis)} }} }
\nc{\Xobsi}{ {\bX_{ i{\rm (obs)} }} }
\nc{\Zobsi}{ {\bZ_{ i{\rm (obs)} }} }
\nc{\bSigobs}{ \bSig_{  {\rm obs} } }
\nc{\bSigmis}{ \bSig_{  {\rm mis} } }
\nc{\bSigmo}{ \bSig_{  {\rm mis,obs} } }
\nc{\bSigom}{ \bSig_{  {\rm obs,mis} } }
\nc{\Xil}{{\bX}_{il}}
\nc{\Zil}{{\bZ}_{il} }
\nc{\omilr}{\omega_{il}^{(r)}}
\nc{\delio}{\bdel_i^{{\rm obs}} }

\nc{\yio}{ {y_i^{\rm o} }}
\nc{\Yio}{ {Y_i^{\rm o }} }
\nc{\Yim}{ {Y_i^{\rm m} }}
\nc{\yim}{ {y_i^{\rm m} }}
\nc{\Yc}{Y^{\rm c}}
\nc{\Yic}{Y_i^{\rm c}}
\nc{\yc}{y^{\rm c}}
\nc{\yic}{y_i^{\rm c}}
\nc{\yi}{y_i}
\nc{\Yi}{Y_i}

\nc{\fyic}{f ( \yic ; \; \psiY )}
\nc{\fyi}{f ( y_i ; \; \psiY ) }
\nc{\fdigivenyic}{f ( d_i  \mid  \yic ; \; \psiDY )}
\nc{\fditilgivenyic}{f ( \tilde{d}_i  \mid  \yic ; \; \psiDY )}
\nc{\fditilgivenyi}{f ( \tilde{d}_i  \mid  \yi ; \; \psiDY )}
\nc{\Fditilgivenyic}{F ( \tilde{d}_i  \mid  \yic ; \; \psiDY )}
\nc{\Fditilgivenyi}{F ( \tilde{d}_i  \mid  \yi ; \; \psiDY )}
\nc{\fdigivenyi}{f (d_i \mid y_i ; \;  \psiDY  )}
\nc{\fyicdi}{f \left( \yic, d_i \right)}
\nc{\fyidi}{f \left( \yi, d_i \right)}

\nc{\fymidr}{f_{Y \mid R}}
\nc{\fyr}{f_{Y,R}}
\nc{\frmidy}{f_{R \mid Y}}
\nc{\fy}{f_Y}
\nc{\fr}{f_R}

\nc{\fyicgivendi}{f (\yic \mid d_i; \; \thetaYD )}
\nc{\fyigivendi}{f (\yi \mid d_i; \; \thetaYD )}
\nc{\fyicgivens}{f (\yic \mid s; \; \thetaYD )}
\nc{\fyigivens}{f (\yi \mid s; \; \thetaYD )}
\nc{\fdi}{f ( d_i; \; \thetaD )}

\nc{\fyicX}{f ( \yic \mid X_i; \; \psiY )}
\nc{\fyiX}{f ( y_i \mid X_i; \; \psiY ) }
\nc{\fdigivenyicX}{f ( d_i  \mid  \yic, X_i ; \; \psiDY )}
\nc{\fdigivenyiX}{f (d_i \mid y_i, X_i ; \;  \psiDY  )}
\nc{\fyicdiX}{f \left( \yic, d_i \mid X_i \right)}

\nc{\fyicgivendiX}{f (\yic \mid d_i, X_i; \; \thetaYD )}
\nc{\fyigivendiX}{f (y_i \mid d_i, X_i; \; \thetaYD )}
\nc{\fdiX}{f ( d_i \mid X_i; \; \thetaD )}

\nc{\Yistar}{\bY_i^*}

\nc{\Dio}{D_i^{\rm obs}}
\nc{\bdelio}{\bdel_{ i \, {\rm (obs)}} }

\nc{\fygivend}{f_{Y \mid \delta}}
\nc{\fyd}{f_{Y, \delta}}
\nc{\fd}{f_\delta}
\nc{\FD}{F_D}
\nc{\fygivenbd}{f_{Y\mid b, \delta}}

\nc{\alphahat}{\hat{\bal}}
\nc{\phihat}{\hat{\bphi}}
\nc{\thetahat}{\hat{\bthe}}
\nc{\thetatilde}{\tilde{\bthe}}
\nc{\scoretheta}{\bS(\bthe; \, \scc)}
\nc{\hesstheta}{\bH(\bthe; \, \scc)}
\nc{\infotheta}{\sci(\bthe; \, \scc)}
\nc{\sitheta}{\bs_i(\bthe; \, \scc_i)}
\nc{\sithetahat}{\bs_i(\thetahat; \, \scc_i)}

\nc{\loglikobs}{\ell_{{\rm o}}(\bthe; \, \sco)}
\nc{\scoreobs}{\bS_{{\rm o}}(\bthe; \, \sco)}
\nc{\hessobs}{\bH_{{\rm o}}(\bthe; \, \sco)}
\nc{\infoobs}{\scj_{{\rm o}}(\bthe; \, \sco)}

\nc{\Cil}{\scc_{il}}
\nc{\olog}{\lambda^*(\bthe, \Xobs)}
\nc{\LthetaC}{\scl(\bthe; \, \scc)}
\nc{\LthetaCi}{\scl_i(\bthe; \, \scc_i)}
\nc{\LthetaCil}{\scl_i (\bthe; \, \scc_{il}) }
\nc{\lthetaC}{\ell(\bthe; \, \scc)}
\nc{\lthetaCi}{\ell_i(\bthe; \, \scc_i)}
\nc{\lthetaCil}{\ell_i (\bthe; \, \scc_{il}) }
\nc{\Qtheta}{\scq \left( \bthe \, \left| \,  \bthe^{(r)} \right. \right)}
\nc{\thetar}{\bthe^{(r)}}
\nc{\thetas}{\bthe^{(s)}}
\nc{\alphas}{\bal^{(s)}}
\nc{\psis}{\psi^{(s)}}
\nc{\alphasplusone}{\bal^{(s+1)}}
\nc{\psisplusone}{\bpsi^{(s+1)}}
\nc{\alphapsis}{\left( \alphas, \psis \right)}

\nc{\thetarplusone}{\bthe^{(r+1)}}
\nc{\ologi}{\lambda^*_i(\bthe, \Xobs)}
\nc{\llogi}{\lambda_i \left( \bthe, \tilde{\Xcom}_{il} \right) }

\nc{\scxil}{\tilde{\Xcom}_{il}}
\nc{\siginv}{\bSig_i^{-1}}

\nc{\fofym}{ f \left( \by_i \mid \bbet_m, \bSig \right) }
\nc{\mphim}{ \phi_M \lsq \bSig^{-1/2}(\by_i - \bX_i \bbet_m) \rsq }
\nc{\mphit}{ \phi_M \lsq \bSig^{-1/2}(\by_i - \bX_i \bbet_{t_i}) \rsq }
\nc{\mphij}{ \phi_M \lsq \bSig^{-1/2}(\by_i - \bX_i \bbet_j) \rsq }
\nc{\mphik}{ \phi_M \lsq \bSig^{-1/2}(\by_i - \bX_i \bbet_k) \rsq }
\nc{\expkerm}{ \exp  \lbc -\half \bu_i(\bbet_m)' \bSig^{-1} \bu_i(\bbet_m)
  \rbc }
\nc{\expkerk}{ \exp  \lbc -\half \bu_i(\bbet_k)' \bSig^{-1} \bu_i(\bbet_k)
  \rbc }
\nc{\expkerj}{ \exp  \lbc -\half \bu_i(\bbet_j)' \bSig^{-1} \bu_i(\bbet_j)
  \rbc }
\nc{\normscorem}{\left( \bX_i' \bSig^{-1} \bX_i \bbet_m - \bX_i' \bSig^{-1}
  \by_i \right) }
\nc{\normscorej}{\left( \bX_i' \bSig^{-1} \bX_i \bbet_j - \bX_i' \bSig^{-1}
  \by_i \right) }
\nc{\piti}{ \pi \left( t_i, \bal, \bZ_i\bgam \right) }
\nc{\omij}{ \om_{ij} \left( t_i, \bal, \bZ_i\bgam \right) }
\nc{\phibetak}{ \phi_M(\bbet_k) }
\nc{\phibetaj}{ \phi_M(\bbet_j) }
\nc{\dphidbetak}{ \left. \dpl \phibetak \right/ \dpl \bbet_k }
\nc{\dphidbetakf}{ \frac{ \dpl \phibetak }{ \dpl \bbet_k } }
\nc{\uik}{\bu_i \left( \bbet_k  \right)}
\nc{\mset}{ \{ 0, 1, \ldots, M \} }
\nc{\betasigma}{ \left( \lbc \bbet^{(r)}_t \rbc, \bSig^{(r)} \right) }
\nc{\Thetar}{ \bThe^{(r)} }

\nc{\shatkm}{\hat{S}_{\rm KM}}

\nc{\ds}{\displaystyle}

\nc{\beq}{\begin{eqnarray*}}
\nc{\eeq}{\end{eqnarray*}}

\nc{\beqna}{\begin{eqnarray}}
\nc{\eeqna}{\end{eqnarray}}

\nc{\bct}{\begin{center}}
\nc{\ect}{\end{center}}

\nc{\bds}{\begin{description}}
\nc{\eds}{\end{description}}

\nc{\bit}{\begin{itemize}}
\nc{\eit}{\end{itemize}}

\nc{\bnu}{\begin{enumerate}}
\nc{\enu}{\end{enumerate}}

\nc{\bgt}{\begin{table}}
\nc{\bgtb}{\begin{center} \begin{tabular}}
\nc{\entb}{\end{tabular} \end{center} }
\nc{\ent}{\end{table}}

\nc{\ts}{\textstyle}

\nc{\bgl}{\begin{letter}}
\nc{\op}{\opening}
\nc{\incl}{\input{sendout.ltr} \closing{Best Regards,} \end{letter} }

\nc{\tb}[1]{\textcolor[rgb]{0.00,0.00,1.00}{#1}}
\nc{\trd}[1]{\textcolor[rgb]{1.00,0.00,0.00}{#1}}
\nc{\tp}[1]{\textcolor[rgb]{1.00,0.00,1.00}{#1}}

\usepackage[title,titletoc,toc]{appendix}
\usepackage{fullpage}
\usepackage{amsmath}
\usepackage{amssymb}
\usepackage{xypic}
\usepackage{lscape}
\usepackage{centernot}
\usepackage{tikz}
\usepackage[affil-it]{authblk}
\usetikzlibrary{arrows}
\usetikzlibrary{fit,positioning}

\usepackage[round]{natbib}

\usepackage[left=2cm,right=2cm,top=3cm,bottom=4cm,]{geometry}

\begin{document}

\title{Joint calibrated estimation of  inverse probability of treatment and censoring weights for marginal structural models}
\author{Sean Yiu\thanks{E-mail address: \texttt{sean.yiu@mrc-bsu.cam.ac.uk}; corresponding author}~  and Li Su}
\affil{MRC Biostatistics Unit, School of Clinical Medicine, University of Cambridge,  Cambridge, CB2 0SR, UK}
\date{}
\maketitle

\baselineskip=24pt
\setlength{\parindent}{.25in}

\begin{center}
{\bf Summary} %\tp{100 words}
\end{center}
Marginal structural models (MSMs) with inverse probability  weighting offer an approach to estimating  causal effects of treatment sequences on repeated outcome measures in the presence of time-varying confounding and dependent censoring. However, when weights are estimated by maximum likelihood,   inverse probability weighted estimators (IPWEs)  can be inefficient and unstable in practice. We propose a joint calibration approach for inverse probability of treatment and censoring weights to  improve the efficiency and robustness of the  IPWEs for MSMs with time-varying treatments of arbitrary (i.e., binary and non-binary)  distributions. Specifically, novel calibration restrictions are derived  by explicitly eliminating covariate associations with both the treatment assignment process and the censoring process after weighting the current sample (i.e., to optimise covariate balance in finite samples). 
A convex minimization procedure is developed  to implement the calibration. Simulations show that IPWEs with calibrated weights perform better than IPWEs  with weights from maximum likelihood. We apply our method to a natural history study of HIV for estimating the cumulative effect of highly active antiretroviral therapy on CD4 cell counts over time. 
\\{\emph{Key words: Calibration; Causal inference; Covariate balance; Dropout; Longitudinal data,  Propensity score.}

%Marginal structural models by inverse probability of treatment weighting (IPTW) offer an approach to estimating the effect of a time-varying treatment on an outcome in the presence of time-dependent confounding. However, if weights are estimated by maximum likelihood, the treatment process after weighting will in general not be statistically exogenous even if the treatment assignment model is correctly specified, due to chance imbalances in the confounder distributions. As a consequence, IPTW can be inefficient. In this article, we calibrate an initial unbiased estimate of the weights by imposing restrictions implying that the treatment process is statistically exogenous after weighting. We extend our method to account for informative censoring, by imposing further restrictions implying that the weighted uncensored observations is a representative sample of the target population based on observed covariates. Simulation studies show that our method is more efficient than the maximum likelihood approach when the treatment assignment model is correctly specified, and is more robust under model misspecification. We apply our method to a natural history study of HIV, where it is of clinical interest to estimate the therapeutic effect of antiretroviral therapy on CD4 cell count. 

\section{Introduction}
\subsection{Drawbacks of the maximum likelihood approach to inverse probability weighting}
Marginal structural models (MSMs) \cite[]{Robins1999a, Robins2000} estimated by inverse probability of treatment weighting (IPTW) are widely used to quantify  causal effects of  treatment sequences on repeated outcome measures in the presence of time-varying confounders that are themselves affected by past treatment  history (e.g., \citealt{Hernan2001, Ko2003}), i.e., in the presence of time-varying confounding \citep{Daniel2013}. For both cross-sectional and longitudinal settings in practice, weights for IPTW are usually obtained by fitting a  parametric  treatment assignment model and then plugging in the parameter estimates from maximum likelihood. 
However, this maximum likelihood approach has important  drawbacks.

First, inverse probability weighted estimators (IPWEs) with weights from  maximum likelihood can be inefficient even when the treatment assignment model  is correctly specified. 
This is because,  weights from maximum likelihood can achieve covariate balance across treatment groups asymptotically, but  are not guaranteed to do so in finite samples, especially when there are many covariates. The situation is analogous to randomised experiments where randomisation balances covariates asymptotically, but not necessarily in finite samples, and substantial imbalance can arise by chance \citep{Pocock2002,Imai2008}. In randomised experiments, it has been shown on numerous occasions that methods that effectively seek to improve covariate balance, e.g., by re-randomisation, blocking, or covariate adjustment, can increase the efficiency of treatment effect estimators \cite[]{Pocock2002, Imai2008,Morgan2012}.  Thus IPWEs with weights that optimise covariate balance in finite samples are likely to be more efficient  than IPWEs with weights from maximum likelihood.

Second, when the treatment assignment model is misspecified,
IPWEs with weights from maximum likelihood can be unstable and have large mean squared error (MSE), even if this misspecification is mild \cite[]{Kang2007, Cole2008, Lefebvre2008}. This is because there is a mismatch between the goals of maximizing the likelihood for predicting treatment assignment  and finding weights that adequately balance covariates. With a slightly   misspecified treatment assignment model, even if the maximised likelihood of this model is large, weighting by the resulting weights from maximum likelihood can still lead to substantial covariate imbalance. To reduce the risk of model misspecification, data-adaptive methods (e.g., machine learning methods) have been used in the literature \cite[]{McCaffrey2013,Gruber2015}. However, the na\"{i}ve use of data-adaptive methods for weight estimation would result in an algorithm that also aims (like maximum likelihood estimation) to achieve optimal prediction of treatment assignment, rather than to optimise covariate balance after weighting.

The second drawback of IPWEs has motivated new covariate balancing weight methods that directly optimise covariate balance for cross-sectional settings (e.g., \citealp{Graham2012, Hainmueller2012, Imai2014, Zubizarreta2015, Chan2016, Fong2018}). These methods have been shown to dramatically improve the performance of IPWEs by reducing MSE under both correct and incorrect model specification. Recent theoretical investigations by \cite{Tan2017} reveal that the improvement brought by the covariate balancing weight methods is because, even  under model misspecification, they can reduce the \emph{relative error} of propensity score (i.e.,  conditional probability of treatment assignment given pre-treatment covariates) estimation, i.e., the ratio of  the true propensity score to its estimated value, which controls the MSE of the IPWE. In contrast, the maximum likelihood approach works on reducing the \emph{absolute error} of propensity score estimation, but this does not necessarily reduce its relative error. For example, when the true propensity scores are small for an area of the covariate space, slight underestimation of these propensity scores can induce large relative error, not absolute error, of propensity score estimation.
In this paper we also show why the maximum likelihood approach may perform poorly in terms of 
 removing covariate imbalances asymptotically  
 under mild model misspecification (see details in Section~\ref{theory} and the Supplementary Material).
%For example, when the true propensity scores for certain covariate space are small, slight underestimation can lead to large  relative error, not  absolute error, of propensity score estimation.

%when the models for the weights are misspecified, IPWe with MLWs can be unstable and have large mean squared error, even if this misspecification is mild [5,16,17]. This is because there is a mismatch between the goals of maximizing the likelihood for predicting treatment assignment (or loss to follow-up) and finding weights that adequately balance covariates [18]. With slight misspecification in the model for treatment assignment (or loss to follow-up), even when the maximised likelihood of this model is large, weighting by the resulting MLWs can lead to substantial covariate imbalance. These phenomena were corroborated by Tan [19] in his theoretical and empirical investigations on the performance of IPWe with MLWs under mild model misspecification.  To reduce the risk of misspecifying the model for the weights, data-adaptive methods (e.g. machine learning methods) have been used in the literature [20,21]. However, the naïve use of data-adaptive methods for weight estimation would result in an algorithm that also aims (like maximum likelihood estimation) to achieve optimal prediction of treatment assignment (or loss to follow-up), rather than to optimise covariate balance after weighting.

Similar to IPTW for MSMs, inverse probability of censoring weighting (IPCW) can be used to address the selection bias due to dependent censoring that is ubiquitous in longitudinal settings \cite[]{Hernan2001}. Within the maximum likelihood framework, this is achieved by fitting a model for the censoring process to obtain another set of time-varying weights. The purpose of  weighting the uncensored observations at a specific time point is to create a representative sample from   the original population without censoring in terms of variables that predict the probability of censoring   at that time point  (e.g., covariate and outcome histories). However, the  two above-mentioned drawbacks of IPWEs also  apply to the case with IPCW \cite[]{Kang2007,Cole2008,Howe2011}. Because in the  maximum likelihood approach, the final weights for fitting MSMs are a product of the time-varying weights for  IPTW and IPCW, these issues are likely to exacerbate when both time-varying confounding and dependent censoring are present.

\subsection{Joint calibration approach to weight estimation}
In this paper we propose methodology to improve the efficiency and robustness of IPWEs   when fitting MSMs with both IPTW and IPCW.   
Our idea is to jointly calibrate  an initial set of time-varying weights (e.g., from maximum likelihood)  for IPTW and IPCW  by imposing covariate balance restrictions simultaneously. Here we use the term `covariate' generally; depending on specific scenarios, it can refer to baseline covariates, time-varying covariates, and history of the repeatedly measured outcome, etc. Specifically, building upon the `covariate association eliminating weights' framework proposed in \cite{Yiu2017} for cross-sectional settings,  we propose  novel calibration restrictions  to explicitly remove  covariate associations over time with both the treatment and censoring processes after weighting  the current sample (i.e., to optimise covariate balance for both treatment assignment and censoring in finite samples). 
  A convex minimization procedure is developed to implement  the joint calibration, where the solution to the restrictions for the calibrated weights is unique and asymptotically equivalent to the initial weights if the models for estimating these initial weights are correctly specified. Thus our calibration procedure maintains the consistency of the IPWEs  with the initial weights. 
 
By enforcing covariate balance as characterized in chosen models for the treatment and censoring processes, our calibrated weights can provide better adjustment for chance imbalances of empirical covariate distributions
than  the maximum likelihood approach  when the models for treatment assignment and censoring are correctly specified, and can be more robust to model misspecification since they are designed to optimise covariate balance. Moreover, our method is applicable to time-varying treatments with arbitrary marginal distributions (e.g., ordinal, categorical and continuous treatments over time), which could greatly promote the flexible and reliable implementation  of MSMs in practice (e.g., the effect of cumulative doses of a treatment on longitudinal outcomes can be estimated).

\subsection{Related methods}
Our research fits into the literature of covariate balancing weights (e.g., \citealp{Graham2012, Hainmueller2012, Imai2014, Zubizarreta2015, Chan2016,Fong2018}), much of which focuses on binary treatments in cross-sectional settings. An exception is the work of \cite{Imai2015} which considers covariate balancing weights with time-varying binary treatments for MSMs in longitudinal settings. However, our method has several characteristics that are distinct from the method by \cite{Imai2015}. First, our method can be applied to  time-varying treatments of arbitrary marginal distributions. In the data example presented in Section~\ref{hersexample}, we focus on ordinal time-varying treatments. Second, we deal with both  time-varying confounding  and dependent censoring that are common in longitudinal settings, while 
\cite{Imai2015} focus on time-varying confounding.  Third, our method can be applied to both  repeated outcome measures over time and an eventual outcome at a study end within an unbalanced observation scheme (i.e., study units can be followed up at different time points), while \cite{Imai2015} deal with an eventual outcome in a balanced observation scheme. Fourth, our method can incorporate  a variety of stabilized weight structures that condition on baseline covariates, while it is not clear how to include arbitrary stabilized weight structures in the  approach by \cite{Imai2015}. Last, the imposed restrictions in our method  do not increase exponentially with the number of time periods unlike in \cite{Imai2015}. Along with the proposed convex minimization procedure, this greatly facilitates the practical implementation of our method, especially when the non-parametric bootstrap is used for making inference.

Recently, \cite{Kallus2018} also proposed a covariate balancing weight approach for binary treatments in longitudinal settings. Specifically, they use kernel smoothing to flexibly model expectations of the potential outcome conditional on treatment and covariate histories up to each time point in the follow-up. Then weights are estimated by minimizing an upper bound for imbalances of time-varying variables (as characterized by conditional expectations of the potential outcome with kernels) over time plus some penalty for the variability of the weights.  Because this approach  uses information from the observed outcome when modeling  conditional expectations of the potential outcome, it is distinct from the standard IPTW approach and our calibrated IPTW approach for MSMs, where  only information for the treatment process is used.  In addition, the  approach in \cite{Kallus2018} involves tuning hyperparameters of the kernels and the penalisation parameter for weight estimation. It is also not clear how to generalize their method to accommodate continuous and other non-binary treatments over time, which is one of the main motivations for developing our method.

%It is also not clear how to generalize it to continuous and other non-binary treatments over time, which are one of the main motivations for our proposed method. 

%The validity of this approach depends upon being able to correctly specify the expectation of the potential outcomes conditional on covariate histories up to each prior visit. Because this is a difficult task owing to the complexity of specifying a compatible set of conditional expectation functions, \cite{Kallus2018} proposed to use kernels and to tune their associated hyper-parameters to approximate the conditional expectation functions. In contrast, the validity of our approach does not depend upon being able to correctly specify these conditional expectation functions, though the efficiency of our IPWE can potentially benefit from their estimation. Instead, our approach is aligned with the standard IPTW approach whose validity depends upon being able to correctly specify the treatment assignment model. This is particularly appealing when the treatment guidelines are well known, e.g., in the motivating application. 

In the context of handling dependent censoring, \cite{Han2016} proposed a calibrated estimation approach for weights in IPCW. We provide a detailed discussion of his approach and compare it with ours for IPCW in Section~\ref{hanswork}. 

\subsection{Motivating example}\label{exampleintro}
Our research is motivated by data from the HIV Epidemiology Research Study (HERS),
 a natural history study of $1310$ women with, or at high risk of, HIV infection at four sites (Baltimore, Detroit, New York, Providence) from 1993 to 2000  \cite[]{Ko2003}. During the study 12 visits were scheduled, where a variety of clinical, behavioural and sociological outcomes as well as self-reported information on antiretroviral therapies  (ARTs) were  recorded  approximately every 6 months. %We will focus on the 871  women who were HIV-positive at enrolment. 

We are interested in quantifying the effect of highly active antiretroviral therapy (HAART),  which contains three or more ART regimens, on the CD4 cell counts over time in the HERS. Because the HERS was an observational study, where therapies were not randomly assigned and varying over time, this leads to the potential for time-varying confounding between treatment and outcome. In particular, important HIV biomarkers such as CD4 cell counts and HIV viral load are affected by previous treatments, but also predict current treatment assignment and subsequent outcome measures over time. Moreover, estimation of the treatment effect may be further complicated by dependent censoring due to patient dropout,  where about half of the 871 HIV-infected women at enrolment did not complete the study.  In the previous analysis by \cite{Ko2003}, weights for IPTW and IPCW were estimated using maximum likelihood to fit several MSMs and address the time-varying confounding and dependent censoring problems in the HERS data. However, the treatment comparison used in \cite{Ko2003} was binary for the groups with `HAART' and `no HAART'. Because patients on  ARTs other than HAART (i.e., less than 3 ARTs) were combined with patients not receiving any treatment, the therapeutic effect of HAART relative to no treatment was likely to be underestimated. In this paper, we consider  the time-varying treatment as ordinal with 3 levels---`no treatment', `ART other than HAART' and `HAART', which  therefore allows more precise quantification of the effect of HAART.  

A key complication of fitting MSMs to the HERS data is the presence of many  time-invariant and time-varying covariates and their interactions, which partly reflects the treatment guideline when the HERS was conducted  \cite[]{Ko2003}. This not only makes it difficult to correctly specify the treatment assignment and censoring models, but suggests that even if this is achieved, IPWEs with weights from  maximum likelihood  might be inefficient because such weights are unlikely to  adequately adjust for chance imbalances of the multivariate covariate distribution. These concerns motivated us to develop more efficient and robust estimators for parameters in  MSMs.

\section{Notation, setting and assumptions}\label{notation}
In this  section, we introduce the notation and  assumptions,  and for clearer exposition, we describe the  setting of interest in the context of the HERS data. 
\subsection{Notation and setting}

In the HERS, the treatment information was self-reported to record the ART use in the last six months prior to the scheduled study visit. Therefore, at  visit $j$ ($j=0, 1,\ldots,T$), we observe, in chronological order, the treatment assignment $A_{ij}$, a vector of time-varying covariates $\bX_{ij}$ (e.g., HIV viral load, HIV symptoms), and the longitudinal outcome $Y_{ij}$ (i.e., CD4 count) from the $i$th patient ($i=1,\ldots,n$). Note that $j=0$ corresponds to baseline and  we allow $A_{ij}$ to be of arbitrary distribution with a possible value $a$. In addition, we  observe $\bV_i$, a vector of baseline covariates such as demographical variables.
In this setting, we assume the temporal ordering where $Y_{ij}$, $\bX_{ij}$ and $A_{ij}$ can only be affected by $\{\overline{A}_{ij},\overline{X}_{ij},\overline{Y}_{i,j-1}, \bV_i\}$, $\{\overline{A}_{ij},\overline{X}_{i,j-1},\overline{Y}_{i,j-1}, \bV_i\}$ and $\{\overline{A}_{i,j-1},\overline{X}_{i,j-1},\overline{Y}_{i,j-1}, \bV_i\}$, respectively, for $j=1, \ldots, T$. Here an overbar represents the history of a process, for example, $\overline{X}_{ij}=\{\bX_{i1},\ldots,\bX_{ij}\}$.  

As mentioned in Section~\ref{exampleintro}, we treat  $A_{ij}$, the ART  use in the previous six months prior to visit $j$ for the $i$th patient, as an ordinal treatment variable. Specifically,  we use two indicator variables $A^0_{ij}$ and $A^1_{ij}$ to represent  $A_{ij}$, where
$A^0_{ij}$ is the indicator of whether at least one ART  was administered, and $A^1_{ij}$ is the indicator of whether HAART was administered given that at least one ART  was administered ($A^0_{ij}=1$).  In particular, `no ART', `one or two ARTs' and `HAART'  are represented by $(0,0)$, $(1,0)$ and $(1,1)$, respectively. We assume that $A_{ij}$ depends on  treatment history $\overline{A}_{i,j-1}$, history of time-varying covariates $\overline{X}_{i,j-1}$, history of longitudinal outcome $\overline{Y}_{i,j-1}$ up to  visit $j-1$ as well as baseline covariates $\bV_i$. This relationship is determined by how the ART information was collected in  the HERS. For other scenarios, appropriate dependence structure in the treatment process can be specified  to reflect the specific contexts; and  our method described below can be easily adapted. For ease of exposition, we absorb $\overline{Y}_{i,j-1}$, $\overline{A}_{i,j-1}$ and $\bV_i$ into the covariate history $\overline{X}_{i,j-1}$ $(j=1,\ldots,T)$, unless stated otherwise.   

Let $Y^{\overline{a}_{j}}_{ij}$  be the potential outcome that would have arisen at visit $j$  had the $i$th patient been assigned the potential treatment sequence  $\overline{a}_{j}$ from the first visit after baseline up to visit $j$. We assume a general MSM of the form $\Ex(Y^{\overline{a}_{j}}_{ij})=\mu(\overline{a}_{j},  \bgamma)=g\{h(\overline{a}_j ), \bgamma\}$, where $h(\cdot)$ is a known function satisfying $h(\overline{a}_{j}=\mathbf{0})=0$, $\mathbf{0}$ is the vector of zeros, $\overline{a}_{j}=\mathbf{0}$ is the potential treatment sequence where no treatment is administered at every visit up to visit $j$,  and $g(\cdot)$ is a known function that relates the mean of the potential outcome to $h(\overline{a}_{j})$ through a finite-dimensional parameter vector $\bgamma$. %For example, if $h(\overline{a}_{j})=\sum_{t=1}^j a_{t}$ and $g\{h(\overline{a}_j), \bV_i,\bgamma\}=\gamma_0+\gamma_1h(\overline{a}_{j})+\bgamma_2^\top\bV_i$, then $\gamma_1$ is the causal effect on $Y^{\overline{a}_{j}}_{ij}$ per unit of cumulative exposure to treatment given $\bV_i$.
%I%n addition, baseline covariates $\bV_i$ can be incorporated into the MSM. 
For example, for the HERS data we may specify $g\{h(\overline{a}_j),\bgamma\}=\gamma_{0j}+\gamma_1\sum_{t=1}^j(a^0_{t}-a^1_{t})+\gamma_2\sum_{t=1}^ja^1_{t}$, where $a^0_{t}$ and $a^1_{t}$ are potential values of treatment indicators $A^0_{it}$ and $A^1_{it}$, and $\gamma_1$ and $\gamma_2$ are the causal effects  on $Y^{\overline{a}_{j}}_{ij}$ per unit increase of cumulative exposures to one or two ARTs and HAART, respectively. %Here baseline covariates $\bV_i$ are incorporated into the MSM to help remove imbalances in these covariates.

\subsection{Assumptions}\label{assumptions}
In order to identify $\bgamma$, MSMs rely on the sequential ignorability of treatment assignment assumption, i.e., $\text{pr}(A_{ij}\mid Y_{ij}^{\overline{a}_{j}},\overline{X}_{i,j-1})=\text{pr}(A_{ij}\mid \overline{X}_{i,j-1})$ for $j=1,\ldots,T$, also known as the assumption of no unmeasured confounders (at each visit/time period). In addition, we make the positivity assumption, i.e., $\text{pr}(A_{ij}\in \mathcal{A}\mid \overline{X}_{i,j-1})>0$ for all $\overline{X}_{i,j-1}$ and for any  set $\mathcal{A}$ with positive measure. Finally,  we make the stable unit treatment value (SUTVA) assumption: the potential outcomes are well defined; the distribution of potential outcomes for one patient is assumed to be independent of potential treatment sequence of another patient.

In the presence of dependent censoring, the objective of MSMs is to  estimate the causal effect of the treatment sequence in the absence of censoring. Let $R_{ij}$ be the indicator of whether the $i$th patient remains in the study up to visit $j$. We assume that $R_{i0}=1$ (i.e., baseline visit assessments are complete for all patients) and  $R_{i,j-1}=0 \Rightarrow R_{ij}=0$ (monotone missingness due to dropout). Our interest is to estimate the parameters of the MSM for $\Ex(Y^{\overline{a}_{j},\overline{r}_{j}=\mathbf{1}}_{ij})$, where $\overline{r}_{j}$ is the potential  sequence of the indicator of the $i$th patient being in the study by visit $j$ and $\mathbf{1}$ is the vector of ones. 
To achieve this, we make an assumption that the censoring is sequentially ignorable, i.e., censoring at visit $j$ depends only on observable history up to but not including visit $j$. Let $\overline{H}_{ij}$ denote this observable history, which may include $\overline{X}_{i,j-1}$, $\overline{A}_{i,j-1}$, $\overline{Y}_{i,j-1}$ and any other relevant covariate information. The sequential ignorability assumption of censoring means that  $\text{pr}(R_{ij}\mid Y_{ij}^{\overline{a}_{j}},\overline{H}_{ij}, R_{i,j-1}=1)=\text{pr}(R_{ij}\mid \overline{H}_{ij}, R_{i,j-1}=1)$ for $j=1,\ldots,T$.
 In addition, we assume that $\text{pr}(R_{ij}\mid \overline{H}_{ij}, R_{i,j-1}=1)>0$ for all $\overline{H}_{ij}$, which is similar to the positivity assumption made for the treatment process. 
 
 Throughout the paper, we make the above assumptions; otherwise our method may result in severely biased  estimates for parameters in the MSM, possibly even compared to   an  analysis without addressing time-varying confounding and dependent censoring.

\section{Inverse probability of treatment weighting}
We first focus on the IPTW approach for dealing with time-varying confounding in MSMs. The IPCW approach for dependent censoring will be described in Section~\ref{IPCW}.

\subsection{Maximum likelihood estimation}

To consistently estimate $\bgamma$, the following estimating equations
\begin{equation}\label{MSM}
\sum_{i=1}^n\sum_{j=1}^{T}SW^A_{ij} D(\overline{A}_{ij},  \bgamma) \left\{Y_{ij}-\mu(\overline{A}_{ij},  \bgamma)\right\}=0
\end{equation} can be solved, 
where  $SW^A_{ij}=\prod_{k=1}^j\text{pr}(A_{ik}\mid \overline{A}_{i,k-1})/\prod_{k=1}^j\text{pr}(A_{ik}\mid \overline{X}_{i,k-1})$ are the stabilized inverse probability of treatment weights (SIPTW),  $D(\overline{A}_{ij}, \bgamma) =\{\partial \mu(\overline{A}_{ij},  \bgamma)/ \partial \bgamma\} V_{ij}^{-1}$ and $V_{ij}=\mbox{var}(Y_{ij})$ \citep{Robins1999a,Hernan2001, Ko2003}. Note that baseline covariates $\bV_i$ can also be included in the numerator of $SW^A_{ij}$ if they are included in the MSM. For simplicity, we do not consider this here, but our method described below easily extends to this scenario.

The intuitive idea behind weighting the $i$th patient's data at visit $j$ by $SW^A_{ij}$ is to create a pseudo-population where $A_{ij}$ does not depend on $\overline{X}_{i,j-1}$ conditional on $\overline{A}_{i,j-1}$,  and  the causal  effect of $\overline{a}_{j}$ on $\Ex(Y^{\overline{a}_{j}}_{ij})$  is the same as in the original population. Under the  sequential ignorablity,   positivity and SUTVA assumptions described in Section~\ref{notation}, the treatment process up to visit $j$ after weighting by $SW^A_{ij}$ will then be causally exogenous \citep{Robins1999a}, i.e.,   $\text{pr}^*(A_{ij}\mid Y_{ij}^{\overline{a}_{j}}, \overline{X}_{i,j-1})=\text{pr}^*(A_{ij}\mid  \overline{X}_{i,j-1})= \text{pr}(A_{ij}\mid \overline{A}_{i,j-1})$, where  $*$ denotes the pseudo-population after weighting by $SW^A_{ij}$.  Therefore, standard regression methods can be used to consistently estimate the parameter $\bgamma$ in the specified MSM if the SIPTW are known.

Because SIPTW are unknown in observational studies, estimates of the SIPTW based on maximum likelihood, $SW^A_{ij}(\hat{\balpha},\hat{\bbeta})=\prod_{k=1}^j\text{pr}(A_{ik}\mid \overline{A}_{i,k-1};\hat{\balpha})/\prod_{k=1}^j\text{pr}(A_{ik}\mid \overline{X}_{i,k-1};\hat{\bbeta})$ are used to implement IPTW, where $\hat{\balpha}$ and $\hat{\bbeta}$ are the maximum likelihood estimates of $\balpha$ and $\bbeta$ in parametric models  $\text{pr}(A_{ij}\mid \overline{A}_{i,j-1};{\balpha})$ and $\text{pr}(A_{ij}\mid \overline{X}_{i,j-1};{\bbeta})$.

\subsection{Calibrated estimation}
\label{CIPTW}
IPTW exploits the fact that $\prod_{k=1}^j\text{pr}^{*}(A_{ik}\mid\overline{X}_{i,k-1})=\prod_{k=1}^j\text{pr}(A_{ik}\mid \overline{A}_{i,k-1})$ for all $j$. However, due to sample randomness, weighting by $SW^A_{ij}(\hat{\balpha},\hat{\bbeta})$ may not remove the associations between $A_{ik}$ and $\overline{X}_{i,k-1}$ conditional on $\overline{A}_{i,k-1}$ for $k=1,\ldots,j$ in finite samples, even if the assumptions in Section~\ref{notation} are satisfied, and the model $\text{pr}(A_{ik}\mid \overline{X}_{i,k-1};\bbeta)$ is correctly specified, i.e., there exists a vector $\bbeta_\text{true}$ such that $\text{pr}(A_{ik}\mid \overline{X}_{i,k-1};\bbeta_\text{true})=\text{pr}(A_{ik}\mid \overline{X}_{i,k-1})$. In other words, there remain chance imbalances and residual confounding of covariates after weighting by $SW^A_{ij}(\hat{\balpha},\hat{\bbeta})$, which lead to finite sample estimation errors \cite[]{Imai2008}.  When $\text{pr}(A_{ik}\mid \overline{X}_{i,k-1};\bbeta)$ is misspecified, weighting by $SW^A_{ij}(\hat{\balpha},\hat{\bbeta})$ may not even guarantee that the associations between treatment assignment and covariates are reduced after weighting relative to the observed data (see the Supplementary Material for more details).

To overcome these problems, our key idea is to  calibrate $SW^A_{ij}(\hat{\balpha},\hat{\bbeta})$ by imposing 
restrictions implying that treatment assignments are unassociated with the history of time-varying covariates over time after weighting the \textit{current} sample.
Specifically, the calibrated weight takes a multiplicative form,
$SW^{A\star}_{ij}(\blambda)=SW^A_{ij}(\hat{\balpha},\hat{\bbeta})c(\overline{X}_{ij},\blambda)$, where $c(\overline{X}_{ij},\blambda)$ is a non-negative function with $c(\overline{X}_{ij},\blambda=\mathbf{0})=1$  and $\blambda$ is a vector of parameters to be estimated. After obtaining  $\hat{\blambda}$, the estimate of $\blambda$ in the calibration procedure (see Section~\ref{Implementation}),  
 we then replace $SW^A_{ij}$ by $SW^{A\star}_{ij}(\hat{\blambda})$  in~\eqref{MSM} to estimate $\bgamma$.

We derive calibration restrictions for SIPTW in MSMs by building on the framework proposed in \cite{Yiu2017} for the cross-sectional setting. Let $\text{pr}(A_{ij}\mid \overline{X}_{i,j-1};\bbeta_\text{w})$ be a parametric model for the treatment assignment.
Here we use the subscript `w' in $\bbeta_\text{w}$ to emphasize that the parametric model used to derive restrictions does not have to be the same as the one used to construct the initial weights $SW^A_{ij}(\hat{\balpha},\hat{\bbeta})$, see Section~\ref{Covariate} for more details.
Following \cite{Yiu2017}, we use the partition 
$\bbeta_\text{w}=\{\bbeta_{\text{wb}},\bbeta_{\text{wd}}\}$, where $\bbeta_\text{wd}$ are the unique parameters that characterize the dependence of $A_{ij}$ on $\overline{X}_{i,j-1}$ excluding the treatment history $\overline{A}_{i,j-1}$ (e.g., regression coefficients of time-varying confounders), and $\bbeta_\text{wb}$ include the intercept terms and parameters that characterize the dependence on treatment history  (e.g., regression coefficients of $\overline{A}_{i,j-1}$). Here the subscripts `d' and `b' stand for dependence and baseline, respectively.
Without loss of generality, let $\text{pr}(A_{ij}\mid \overline{X}_{i,j-1};\bbeta_{\text{wb}}=\balpha,\bbeta_{\text{wd}}=\mathbf{0})=\text{pr}(A_{ij}\mid \overline{A}_{i,j-1};\balpha)$, i.e., setting $\{\bbeta_{\text{wb}}=\balpha,\bbeta_{\text{wd}}=\mathbf{0}\}$ results in a treatment process model that only depends on treatment history and is parameterized by $\balpha$.  

Now suppose that $\blambda$ is fixed and we have  known weights $SW^{A\star}_{ij}(\blambda)$, it is possible to check whether $\bbeta_{\text{wd}}=\mathbf{0}$ in the pseudo-population over time after weighting the current sample with $SW^{A\star}_{ij}(\blambda)$,  by finding the value of $\bbeta_{\text{w}}$ that maximizes
\begin{equation}\label{treatlike}
\prod_{j=1}^T \prod_{i=1}^n\left\{\prod_{k=1}^j\text{pr}(A_{ik}\mid \overline{X}_{i,k-1};\bbeta_\text{w})\right\}^{SW^{A\star}_{ij}({\footnotesize \blambda})}, 
\end{equation}
or solves the score equations 
\begin{equation}\label{treatscore}
\sum_{j=1}^T \sum_{i=1}^n SW^{A\star}_{ij}(\blambda)\sum_{k=1}^j\frac{\partial}{\partial \bbeta_\text{w}} \log\{\text{pr}(A_{ik}\mid \overline{X}_{i,k-1};\bbeta_\text{w})\}=0.
\end{equation}
The terms in the first product in \eqref{treatlike} makes it explicit that we would like to use $SW^{A\star}_{ij}(\blambda)$ to weight the likelihood of the observed treatment sequence  for the $i$th patient up to visit $j$. \eqref{treatlike} is then constructed by aggregating these terms over all patients and visits for which we require weights $SW^{A\star}_{ij}(\blambda)$. 

%For full generality, we have not imposed a particular structure on $\bbeta_{\text{w}}$ since in practice it can take a variety of forms, e.g., $\bbeta_{\text{w}}$ could be time-invariant, time-dependent but with a parametric structure, or even visit-specific with $\bbeta_{\text{w}}^j$ ($j=1, \ldots, T$).

We propose to derive calibration restrictions by inverting~\eqref{treatscore}, so that we are finding the value of $\blambda$ implying that $\{\bbeta_{\text{wb}}=\hat{\balpha},\bbeta_{\text{wd}}=\mathbf{0}\}$ are the values that maximize~\eqref{treatlike}. That is, we solve for $\blambda$
\begin{equation}\label{treatrestrict}
\sum_{j=1}^T \sum_{i=1}^n SW^{A\star}_{ij}({\blambda})\sum_{k=1}^j\frac{\partial}{\partial \bbeta_\text{w}} \log\{\text{pr}(A_{ik}\mid \overline{X}_{i,k-1};\bbeta_\text{w})\}\Big|_{{\left\{{\footnotesize \bbeta}_{\text{wb}}=\hat{\footnotesize\balpha},{\footnotesize \bbeta}_{\text{wd}}=\mathbf{0}\right\}}}=0.
\end{equation}
%That is, after weighting by $SW^{A\star}_{ij}(\hat{\blambda})$ with  ~\eqref{treatrestrict} satisfied, $\{\bbeta_{\text{wb}}=\hat{\balpha},\bbeta_{\text{wd}}=\mathbf{0}\}$ are the values that maximize~\eqref{treatlike}. %Therefore,   $\prod_{i=1}^n\prod_{k=1}^j\text{pr}^*(A_{i,k-1}\mid \overline{X}_{ik-1};\hat{\bbeta}_\text{w})=\prod_{i=1}^n\prod_{k=1}^j\text{pr}(A_{i,k-1}\mid \overline{A}_{ik-1};\hat{\balpha})$,  \tp{ should this be aggregated over time and across patients?} 
Satisfaction of the restrictions in~\eqref{treatrestrict}  means that after weighting by $SW^{A\star}_{ij}(\hat{\blambda})$ the treatment assignments up to visit $j$ are unassociated with the histories of the time-varying covariates conditional on the treatment histories in the current sample (i.e., $\bbeta_{\text{wd}}=\mathbf{0}$). Note that the structure of the covariate associations is characterized by the specified  parametric  treatment process model. More discussion about this general framework for weight estimation in cross-sectional settings  can be found in \cite{Yiu2017}.
 
Another property of the true SIPTW is $\Ex(SW^A_{ij})=1$  for $j=1,\ldots,T$, where the expectation is taken with respect to $\{\overline{A}_{ij}, \overline{X}_{i,j-1}, \overline{Y}_{i,j-1},\bV_i\}$ \cite[]{Cole2008}. However, in practice the average of the estimated weights by maximum likelihood $SW^A_{ij}(\hat{\balpha},\hat{\bbeta})$ at each visit can take values very different from one, particularly if the treatment process model is badly misspecified.  In order to help stabilize the weights, we  propose to further impose the restrictions
\begin{equation}\label{normalizerestrict}
\frac{1}{n}\sum_{i=1}^n SW^{A\star}_{ij}({\blambda})=1
\end{equation}
for  $j=1,\ldots,T$, in the same spirit as in  \cite{Cao2009}. That is, we constrain the average of the weights to be one at each visit.  The restriction in~\eqref{normalizerestrict} also prevents the trivial solution of zeros for the weights  in~\eqref{treatrestrict} when  only IPTW is applied (see Section~\ref{combiningrestrictions} for more details). %\tp{\eqref{normalizerestrict} is also required for arbitrary calibration functions in order to prevent the trivial solution of zeros for the weights in~\eqref{treatrestrict} if only IPTW is used, see Section~\ref{combiningrestrictions} for more details.}

For the setting with an eventual outcome  at visit $T$ (e.g., the CD4 count at the study end of the HERS), restrictions can be derived by using the above procedure. However,  since in this case we would only be interested in calibrating $SW^A_{iT}(\hat{\balpha},\hat{\bbeta})$, ~\eqref{treatlike} will only contain the terms weighted by $SW^{A\star}_{iT}({\blambda})$, i.e., the terms where $j=T$ in~\eqref{treatlike}. This results in the restrictions
\begin{equation*}
\sum_{i=1}^n SW^{A\star}_{iT}({\blambda})\sum_{k=1}^T\frac{\partial}{\partial \bbeta_\text{w}} \log\{\text{pr}(A_{ik}\mid \overline{X}_{i,k-1};\bbeta_\text{w})\}\Big|_{{\left\{{\footnotesize \bbeta}_{\text{wb}}=\hat{\footnotesize\balpha},{\footnotesize \bbeta}_{\text{wd}}=\mathbf{0}\right\}}}=0~~\text{and}~~\frac{1}{n}\sum_{i=1}^n SW^{A\star}_{iT}({\blambda})=1.
\end{equation*}

\subsection{Application to time-varying ordinal treatment}
We consider the following model for the ordinal treatment variable in the HERS data,  
\begin{equation}\label{ordinaltreatment}
\begin{split}
&\logit \{\text{pr}(A_{ij}^0=1\mid \overline{X}_{i,j-1})\}=\widetilde{\bX}_{i,j-1}^{0\top}\bbeta^{0},\\
&\logit \{\text{pr}(A_{ij}^1=1\mid  \overline{X}_{i,j-1},A_{ij}^0=1)\}=\widetilde{\bX}_{i,j-1}^{1\top}\bbeta^{1},
\end{split}
\end{equation}
where $\widetilde{\bX}_{i,j-1}^0$ and $\widetilde{\bX}_{i,j-1}^1$ include $1$ and functionals of  $\overline{X}_{i,j-1}$ (e.g., transformations and interactions), and 
$\bbeta^{0}$ and $\bbeta^{1}$ are corresponding regression coefficients. 
%\text{pr}(A_{ij}\mid \overline{X}_{i,j-1};\beta_{A^0},\beta_{A^1})&=\text{expit}\{(2A^0_{ij}-1)\beta_{A^0}^\top[1,\overline{X}_{i,j-1}]\}\\
%&\times\left[\text{expit}\{(2A^1_{ij}-1)\beta_{A^1}^\top[1,\overline{X}_{i,j-1}]\}\right]^{A^0_{ij}}
%^\top[1,\overline{X}_{i,j-1}
%where $\text{expit}(x)=1/\{1+\exp(-x)\}$. This model can be fitted by applying logistic regression to $A^0_{ij}$ and then to $A^1_{ij}$ for those with $A^0_{ij}=1$. 
Following Section~\ref{CIPTW}, restrictions based on~\eqref{ordinaltreatment} can  be derived:
\begin{equation}\label{ordinaltreatmentrestric}
\begin{split}
\sum_{j=1}^T \sum_{i=1}^nSW^{A\star}_{ij}({\blambda})\sum_{k=1}^j\left(A^0_{ik}-\hat{e}_{ik}^0\right)\widetilde{\bX}_{i,k-1}^0&=0,\\
\sum_{j=1}^T\sum_{i=1}^nSW^{A\star}_{ij}({\blambda})\sum_{k=1}^jA^0_{ik}\left(A^1_{ik}-\hat{e}_{ik}^1\right)\widetilde{\bX}_{i,k-1}^1&=0,\\
\end{split}
\end{equation}
where $\hat{e}_{ik}^0$ and $\hat{e}_{ik}^1$ are the predicted probabilities of receiving treatment at visit $k$ from fitting the model~\eqref{ordinaltreatment} but with treatment history as the only covariates. The restrictions in~\eqref{ordinaltreatmentrestric} are in spirit similar to the covariate balancing restrictions/conditions for binary treatments in cross-sectional settings  \cite[]{Imai2014,Yiu2017}, but they are aggregated over time. Examining these restrictions carefully, we can see that they aim to remove the associations of the covariates, $\tilde{\bX}_{i,j-1}^0$ and  $\tilde{\bX}_{i,j-1}^1$, with  the residuals of the treatment variables (after fitting \eqref{ordinaltreatment} with  treatment history as the only covariates) in the pseudo-population over time.  Without the general framework described in Section~\ref{CIPTW}, it is not obvious how to generalize the restrictions in cross-sectional settings (e.g., in  \cite{Imai2014}) to longitudinal settings and to arbitrary treatment distributions. 

As a further example, we derive restrictions for treatment sequences with continuous marginal distributions  in the Supplementary Material.

\section{Inverse probability of censoring weighting}\label{IPCW}

%Censoring often occurs in many longitudinal studies, e.g., the HERS. This adds a further complication because the task at hand will generally be to estimate the causal effect of treatment in the absence of censoring. More formally, we would like to estimate the parameters of the marginal structural model $E(Y^{\overline{a}_{ij},\overline{c}_{ij}=1}_{ij}\mid V_{i1})-E(Y^{0,\overline{c}_{ij}=1}_{ij}\mid V_{i1})=\gamma^\top h(\overline{a}_{ij})$ where $\overline{c}_{ij}=1$ is the potential censoring sequence of the $i$th unit being in the study at time $j$. 
\subsection{Maximum likelihood estimation}

Recall that in the presence of censoring, our interest is to 
 estimate the parameters of the MSM for $\Ex(Y^{\overline{a}_{j},\overline{r}_{j}=\mathbf{1}}_{ij})$.
If $R_{ij}$ only depends on the treatment history $\overline{A}_{i,j-1}$ and the MSM for $\Ex(Y^{\overline{a}_{j},\overline{r}_{j}=\mathbf{1}}_{ij})$ is correctly specified, we can still  consistently estimate the parameters in this new MSM with  IPTW,  but  the summands in~\eqref{MSM} are multiplied by $R_{ij}$. In the calibration approach, we would replace $SW^{A\star}_{ij}({\blambda})$ with $R_{ij}SW^{A\star}_{ij}({\blambda})$ in~\eqref{treatrestrict} and alter the scaling in~\eqref{normalizerestrict} so that the average of the weights is still fixed at one for each visit (i.e., replace $n$ by $\sum_{i=1}^n R_{ij}$ at visit $j$).

In the presence of dependent censoring, where $R_{ij}$ also depends on $\overline{X}_{i,j-1}$ including the outcome history $\overline{Y}_{i,j-1}$, conditioning on uncensored observations for analysis will induce  selection bias   when estimating the parameters in the MSM because $\text{pr}(\bX_{i,k-1}\mid \overline{A}_{i,k-1},R_{ij}=1)\neq \text{pr}(\bX_{i,k-1}\mid \overline{A}_{i,k-1})$ $ (k=1,\ldots,j)$.  For example, in the HERS the probability of dropout  at the current visit might depend on most recent changes in the CD4 count outcome \cite[]{Ko2003}. 

%and $\overline{X}_{i,j-1}$ affects $\Ex(Y^{\overline{a}_{j},\overline{r}_{j}=\mathbf{1}}_{ij})$, then IPTW will in general produce biased estimates of the parameters in the new MSM because $\Ex(\bX_{ik-1}\mid \overline{A}_{ik-1},R_{i,k-1}=1)\neq \Ex(\bX_{ik-1}\mid \overline{A}_{ik-1})$ $(k=1,\ldots,j)$, i.e., there is selection bias due to dependent censoring. 

Under the sequential ignorability and positivity assumptions for censoring described in Section~\ref{assumptions},  
we can apply inverse probability of treatment and censoring weighting (IPTCW) and   replace the weights in~\eqref{MSM} by $R_{ij}SW^A_{ij}W^C_{ij}$,  where $W^C_{ij}=\prod_{k=1}^j1/\text{pr}(R_{ik}=1\mid \overline{H}_{i,k-1},R_{i,k-1}=1)$ ($j=1, \ldots, T$) are the inverse probability of censoring weights.
$W^C_{ij}$ are typically estimated by maximum likelihood after specifying a parametric model $\text{pr}(R_{ik}=1\mid \overline{H}_{i,k-1},R_{i,k-1}=1; {\btheta})$; and  $W^C_{ij}(\hat{\btheta})=\prod_{k=1}^j1/\text{pr}(R_{ik}=1\mid \overline{H}_{i,k-1},R_{i,k-1}=1;\hat{\btheta})$, where $\hat{\btheta}$ are  maximum likelihood estimates of $\btheta$.  

Similarly to IPTW,  stabilized weights for censoring, $SW^C_{ij}(\hat{\bet},\hat{\btheta})=W^C_{ij}(\hat{\btheta})\prod_{k=1}^j\pi^{\text{s}}_{ik}(\hat{\bet})$, can be used, where $\pi^{\text{s}}_{ik}(\bet)=\text{pr}(R_{ik}=1\mid \overline{A}_{i,k-1},R_{i,k-1}=1;{\bet})$ is a parametric model for the censoring process given the treatment history only. In the next section, we describe our proposed method for calibrating unstabilized weights for censoring; details about calibrating stabilized weights for censoring can be found in the Supplementary Material.

\subsection{Calibrated estimation}\label{censorcal}

\subsubsection{The two time period setting}\label{cross-sectional}
We consider the two time period setting (including baseline, i.e., $T=1$) to convey the idea as to how the censoring weights $W^C_{ij}(\hat{\btheta})$ can be calibrated by covariate balancing. Since $R_{i0}=1~(i=1,\ldots,n)$, the censoring only occurs at the first follow-up visit in the two time period setting. %For simpler exposition, we therefore drop the subscript for  the second visit  in this section. 
% to have $W^{C\star}_{ij}(\blambda)=W^C_{i}(\hat{\btheta})R_{i}(\lambda)$, to increase efficiency.
The aim of weighting the complete cases (those with $R_{i1}=1$) by $1/\pi_{i1}$, where $\pi_{i1}=\text{pr}(R_{i1}=1\mid \overline{H}_{i0})$ is the true probability of remaining in the study at the first follow-up visit, is to create a representative sample of the target population (i.e., the population that would have been observed in the absence of censoring) in terms of the covariates $\overline{H}_{i0}$. 
Specifically, if the complete cases receive a weight of \textit{one }to represent themselves, the additional weights $1/\pi_{i1}-1$ given to the complete cases are used to create a representative sample of the \textit{incomplete} cases (those with $R_{i1}=0$).  As a result, the total $\sum_{i=1}^n 1/\pi_{i1}$ copies of the complete cases form the pseudo-population after weighting, which can represent  the target population. 
%This weighting approach achieves this because weighting the complete cases by $1/\pi_i-1$ creates a representative sample of the incomplete cases (those with $C_i=0$) and thus additionally weighting the complete cases once (to represent themselves) results in a representative sample of the target population.

%In particular, IPCW exploits the fact that weighting the complete cases by $1/\pi_{i1}-1$ and the incomplete cases by one results in $\text{pr}^{*}(R_{i1}=1\mid \overline{H}_{i0})=1/2= 1-\text{pr}^{*}(R_{i1}=1\mid \overline{H}_{i0})$, where $*$ denotes the pseudo-population after weighting only the complete cases by $1/\pi_{i1}-1$ (see the proof  in the Supplementary Material). 

Let $*$ denote the pseudo-population after weighting only the complete cases by $1/\pi_{i1}-1$. IPCW exploits the fact that $\text{pr}^{*}(R_{i1}=1\mid \overline{H}_{i0})=1/2= 1-\text{pr}^{*}(R_{i1}=1\mid \overline{H}_{i0})$ (see the proof in the Supplementary Material). However, in finite samples, due to sample randomness, weighting the complete cases by $W^C_{i1}(\hat{\btheta})-1$ will not necessarily represent the incomplete cases in terms of covariate distributions. This motivates our calibration approach for the weights in IPCW.

Using the multiplicative form as in Section~\ref{CIPTW}, we can write the calibrated weights as $W^{C\star}_{i1}(\blambda)=W^C_{i1}(\hat{\btheta})c(\overline{H}_{i0},\blambda)$, where $c(\overline{H}_{i0},\blambda)$ is a non-negative function satisfying $c(\overline{H}_{i0},\blambda=\mathbf{0})=1$, and $\blambda$ is the parameter vector to be estimated for the calibration.  Let $\pi_{i1}(\btheta_\text{w})=\text{pr}(R_{i1}=1\mid \overline{H}_{i0}; \btheta_\text{w}) $ be a parametric model for the censoring process with a parameter vector $\btheta_\text{w}$. Now fixing $\blambda$ and weighting the complete cases with the known weights $W^{C\star}_{i1}(\blambda)-1$ to represent the incomplete cases, we can 
 construct the  likelihood for the pseudo-population  excluding the copies of themselves for the complete cases as
\[\prod_{i=1}^n\pi_{i1}(\btheta_\text{w})^{R_{i1}\left\{W^{C\star}_{i1}\left({\footnotesize \blambda}\right)-1\right\}}\left\{1-\pi_{i1}(\btheta_\text{w})\right\}^{1-R_{i1}}.\]
The corresponding score equations are
\begin{equation}\label{missingscore}
\sum_{i=1}^nR_{i1}\left\{W^{C\star}_{i1}(\blambda)-1\right\}\frac{\partial}{\partial \btheta_\text{w}}\log\{\pi_{i1}(\btheta_\text{w})\}+(1-R_{i1})\frac{\partial}{\partial \btheta_\text{w}}\log\{1-\pi_{i1}(\btheta_\text{w})\}=0.
\end{equation}
Solving~\eqref{missingscore} in terms of $\btheta_\text{w}$ provides a measure of how the additional $W^{C\star}_{i1}(\blambda)-1$ copies of complete cases differ from the incomplete cases. In particular, without loss of generality, let $\pi_{i1}(\btheta_{\text{w}}=\mathbf{0})=1/2$.  Then if  a solution to~\eqref{missingscore} is $\hat{\btheta}_w=\mathbf{0}$, this would suggest that the $W^{C\star}_{i1}(\blambda)-1$ copies of complete cases are representative of the incomplete cases in terms of $\overline{H}_{i0}$, while deviations from zero suggest otherwise. We propose to derive the restrictions for calibration by finding $\blambda$ such that $\btheta_{\text{w}}=\mathbf{0}$ are the values that solve \eqref{missingscore}. That is, we solve for $\blambda$ such that
\begin{equation}\label{invertmissingscore}
\sum_{i=1}^nR_{i1}\left\{W^{C\star}_{i1}(\blambda)-1\right\}\frac{\partial}{\partial \btheta_\text{w}}\log\{\pi_{i1}(\btheta_\text{w})\}+(1-R_{i1})\frac{\partial}{\partial \btheta_\text{w}}\log\{1-\pi_{i1}(\btheta_\text{w})\}\Big|_{{\footnotesize\btheta}_{\text{w}}=\mathbf{0}}=0.
\end{equation}
 For example,  with a logistic model  $\logit\{\pi_{i1}(\btheta_{\text{w}})\}=\widetilde{\bH}_{i0}^{\top}\btheta_{\text{w}}$, where $\widetilde{\bH}_{i0}$ is a vector of functionals of $\overline{H}_{i0}$ including $1$, the restrictions based on \eqref{invertmissingscore} are 
\begin{eqnarray}
&\sum_{i=1}^n\{R_{i1}W^{C\star}_{i1}({\blambda})-1\}\widetilde{\bH}_{i0}=0  \nn\\
&\equiv \sum_{i=1}^n R_{i1}W^{C\star}_{i1}({\blambda})\widetilde{\bH}_{i0}=\sum_{i=1}^n\widetilde{\bH}_{i0}. \label{censorbinary}
\end{eqnarray}
These restrictions constrain the sample size after weighting  to be $n$ (since $\widetilde{\bH}_{i0}$ includes 1) and the weighted averages of other elements of $\widetilde{\bH}_{i0}$ to be equal to their averages in the observed data. Although motivated differently, these restrictions have been considered for weight estimation (e.g., \citealt{Robins2007,Vansteelandt2012,Zubizarreta2015}, among others). 

%Restrictions for the stabilized weights  can be derived, e.g.,  by multiplying the summands in~\eqref{censorbinary} by $\text{pr}(R_{i1}=1\mid A_{i0};\hat{\bet})$, 
%$$ \sum_{i=1}^n R_{i1}SW^{C\star}_{i1}({\blambda})\widetilde{\bH}_{i0}=\sum_{i=1}^n\text{pr}(R_{i1}=1\mid A_{i0};\hat{\bet})\widetilde{\bH}_{i0}, $$ where
%$SW^{C\star}_{i1}({\blambda})=W^C_{ij}(\hat{\btheta})\text{pr}(R_{i1}=1\mid A_{i0};\hat{\bet})c_{i1}(\blambda)$. 

%where  $\text{pr}(R_{i1}=1\mid \bV_{i};{\bet})$ is a parametric model for the censoring process given baseline covariates with a parameter vector ${\bet}$,  and  $\hat{\bet}$ is  the corresponding maximum likelihood estimate. %From these restrictions, it is clear that $\text{pr}(C_i=1\mid A_i;\hat{\eta})$ can help reduce the variability of the weights, particularly when it predicts $C_i$ well.

\subsubsection{The general longitudinal setting}
   %, with the convention that $\text{pr}(R_{i,k-1}=1\mid \overline{H}_{ik-1},R_{ik-1}=0)=0$ and $R_{i0}=1$ for all $i$. 
The purpose of weighting the complete cases at visit $j$ (those with $R_{ij}=1$) is to create a representative sample of the target population (in the absence of  censoring) at visit $j$. Let  $\pi_{ik}=\text{pr}(R_{ik}=1\mid \overline{H}_{i,k-1},R_{i,k-1}=1)$  be the true conditional probability of remaining in the study at  visit $k$ given the covariate history $\overline{H}_{i,k-1}$ and that the patient is still under follow-up at visit $k-1$ ($k=1, \ldots, j$). We can show by induction that the weights $W^C_{ij}=1/\prod_{k=1}^{j}\pi_{ik}$ can be used to achieve this purpose.
Specifically, we know from Section~\ref{cross-sectional} that weighting the complete cases at the first follow-up visit by $W^C_{i1}$ will create a representative sample of the target population when $j=1$, so the base case holds. Now we assume that the proposition holds at visit $k-1$, i.e., weighting the complete cases at visit $k-1$ by $W^C_{i,k-1}$ creates a representative sample of the target population at visit $k-1$; and  we treat this weighted population at visit $k-1$ as our new population. The inductive step  at visit $k$ requires showing that weighting the complete cases of this new population at visit $k$ by $1/\pi_{ik}$ will create a representative sample of this new population if no censoring occurs. Using the same logic in Section~\ref{cross-sectional}, it is easy to show that this inductive step holds for the true weight $1/\pi_{ik}$.
By the proposition at visit $k-1$, we will then have a representative sample of the target population at visit $k$ by weighting with $W^C_{ik}$.

We derive calibration restrictions  by following the same strategy as in Section~\ref{cross-sectional}. First, we fix $\blambda$ so that we have known weights $W^{C\star}_{i1}(\blambda),\ldots,W^{C\star}_{ij}(\blambda)$, where $W^{C\star}_{ik}(\blambda)=W^C_{ik}(\hat{\btheta})c(\overline{H}_{i,k-1},\blambda)$ and $c(\overline{H}_{i,k-1},\blambda)$ is a non-negative function satisfying $c(\overline{H}_{i,k-1},\blambda=\mathbf{0})=1$ for $k=1, \ldots, j$.  We can check the validity of the proposition at  visit $j$ by specifying a parametric model $\pi_{ik}(\btheta_\text{w})=\text{pr}(R_{ik}=1\mid \overline{H}_{i,k-1},R_{i,k-1}=1;\btheta_\text{w})$ $(k=1,\ldots,j)$ and estimating its parameter $\btheta_\text{w}$ by maximizing 
\begin{equation}\label{missingliketimej}
\prod_{i=1}^n\prod_{k=1}^j\left[\pi_{ik}(\btheta_\text{w})^{R_{ik}\left\{1/\pi^\star_{ik}({\scriptsize\blambda})-1\right\}}\left\{1-\pi_{ik}(\btheta_\text{w})\right\}^{1-R_{ik}}\right]^{W^{C\star}_{i,k-1}({\scriptsize \blambda})R_{i,k-1}},
\end{equation}
 where $1/\pi^\star_{ik}(\blambda)=W^{C\star}_{ik}(\blambda)/W^{C\star}_{i,k-1}(\blambda)$ $(k=1,\ldots,j)$,   $W^{C\star}_{i0}(\blambda)=1$ and by convention  $0^0=1$. 
The terms in~\eqref{missingliketimej} are used to check the validity of the inductive steps at times $k=1,\ldots,j$ assuming the proposition holds at visit $k-1$. In particular, deviations from ${\btheta}_\text{w}=\mathbf{0}$ provides evidence against the inductive step at one or more visits up to and including visit $j$, and thus evidence against the proposition at visit $j$. Similarly, we can simultaneously check the validity of the proposition at each visit  by maximizing 
 \begin{equation}\label{missingliketime}
\prod_{j=1}^T \prod_{i=1}^n\prod_{k=1}^j\left[\pi_{ik}({\btheta}_\text{w})^{R_{ik}\left\{1/\pi^\star_{ik}({\scriptsize\blambda})-1\right\}}\left\{1-\pi_{ik}({\btheta}_\text{w})\right\}^{1-R_{ik}}\right]^{W^{C\star}_{i,k-1}({\scriptsize \blambda})R_{i,k-1}}, 
\end{equation} which is obtained by aggregating~\eqref{missingliketimej} across $j=1,\ldots,T$.
We can further simplify \eqref{missingliketime} to
\begin{equation}\label{missingliketimesim}
\prod_{j=1}^T \prod_{i=1}^n \left[\pi_{ij}({\btheta}_\text{w})^{R_{ij}\left\{1/\pi^\star_{ij}({\scriptsize\blambda})-1\right\}}\left\{1-\pi_{ij}({\btheta}_\text{w})\right\}^{1-R_{ij}}\right]^{(T-j+1)W^{C\star}_{i,j-1}({\scriptsize \blambda})R_{i,j-1}}
\end{equation}
with the score equations
\begin{equation}\label{missingscorelong}
\begin{split}
&\sum_{j=1}^T \sum_{i=1}^n(T-j+1) \left[  R_{ij}\left\{W^{C\star}_{ij}(\blambda)-W^{C\star}_{i,j-1}(\blambda)\right\}\frac{\partial}{\partial \btheta_\text{w}}\log \left\{\pi_{ij}(\btheta_\text{w})\right\} \right.\\
+& \left.W^{C\star}_{i,j-1}(\blambda)(R_{i,j-1}-R_{ij})\frac{\partial}{\partial \btheta_\text{w}}\log\{1-\pi_{ij}(\btheta_\text{w})\} \right]=0.
\end{split}
\end{equation}
The inductive steps in~\eqref{missingliketimesim} are weighted  by $T-j+1$ to reflect that they are required for checking $T-j+1$ propositions, specifically if the proposition holds at visits $j,\ldots,T$. We derive restrictions by finding $\blambda$ such that $\btheta_\text{w}=\mathbf{0}$ are the values that solve~\eqref{missingscorelong}. That is, we solve for $\blambda$ such that
\begin{equation}\label{generalmissingrestrict}
\begin{split}
&\sum_{j=1}^T \sum_{i=1}^n(T-j+1) \left[  R_{ij}\left\{W^{C\star}_{ij}(\blambda)-W^{C\star}_{i,j-1}(\blambda)\right\}\frac{\partial}{\partial \btheta_\text{w}}\log \left\{\pi_{ij}(\btheta_\text{w})\right\} \right.\\
+& \left.W^{C\star}_{i,j-1}(\blambda)(R_{i,j-1}-R_{ij})\frac{\partial}{\partial \btheta_\text{w}}\log\{1-\pi_{ij}(\btheta_\text{w})\} \right]\Big|_{{\scriptsize\btheta}_\text{w}=\mathbf{0}}=0.
\end{split}
\end{equation}

In this paper we assume a logistic model  $\logit\{\pi_{ij}(\btheta_{\text{w}})\}=\widetilde{\bH}_{i,j-1}^{\top}\btheta_{\text{w}}$, where $\widetilde{\bH}_{i, j-1}$ is a vector of   functionals of $\overline{H}_{i,j-1}$ including $1$.  
The restrictions based on~\eqref{generalmissingrestrict} are 
\begin{eqnarray}
&&\sum_{j=1}^T(T-j+1)\sum_{i=1}^n\left[R_{ij}W^{C\star}_{ij}({\blambda})-R_{i,j-1}W^{C\star}_{i,j-1}({\blambda}) \right] \widetilde{\bH}_{i, j-1}=0.  \label{missingrestrict}
\end{eqnarray}  The term $\sum_{i=1}^n\left[R_{ij}W^{C\star}_{ij}({\blambda})-R_{i,j-1}W^{C\star}_{i,j-1}({\blambda}) \right] \widetilde{\bH}_{i, j-1}$
in~\eqref{missingrestrict} can be interpreted as  the balance summary of $ \widetilde{\bH}_{i, j-1}$  between the weighted  complete cases at visit $j$ and the weighted complete cases at visit $j-1$. 

The restrictions in~\eqref{missingrestrict} are equivalent to 
\begin{eqnarray} \label{missingrestrict2}
\sum_{j=1}^T \sum_{i=1}^n R_{ij}W^{C\star}_{ij}({\blambda})\left\{(T-j+1)\widetilde{\bH}_{i, j-1}-(T-j)\widetilde{\bH}_{ij}\right\}=T\sum_{i=1}^n\widetilde{\bH}_{i0};
\end{eqnarray}  details can be found in the Supplementary Material.
Since $\widetilde{\bH}_{i, j-1}$ ($j=1, \ldots, T$)  includes 1, then 
\eqref{missingrestrict2}  imposes 
$$
\sum_{j=1}^T\sum_{i=1}^n R_{ij}W^{C\star}_{ij}({\blambda})=nT, $$
which means that the total number of `observations' after weighting is equal to $nT$,  the total number of observations of the target population if no censoring occurs at all. 
If $\widetilde{\bH}_{i, j-1}$   includes baseline covariates $\bV_i$, \eqref{missingrestrict2} imposes $$\sum_{j=1}^T\sum_{i=1}^n R_{ij}W^{C\star}_{ij}({\blambda}) \bV_i=T\sum_{i=1}^n \bV_{i}, $$ i.e., the weighted average of $\bV_i$ over all visits is equal to the sample average of $\bV_i$. 
If $\widetilde{\bH}_{i, j-1}$ includes  an indicator for visit, ${I}(j=k)$ ($k=1,\ldots,T$), and  an interaction between this visit indicator and $\bV_i$, ${I}(j=k)\bV_i$, then \eqref{missingrestrict2} imposes $$\sum_{i=1}^nR_{ik}W^{C\star}_{ik}({\blambda})=n, $$ $$\sum_{i=1}^nR_{ik}W^{C\star}_{ik}({\blambda})\bV_i=\sum_{i=1}^n \bV_i $$  for $k=1,\ldots,T$, i.e., at each visit the sample size after weighting is $n$ and the weighted average of $\bV_i$ is equal to the sample average of $\bV_i$.

%Restrictions for stabilized weights can be derived similarly as in Section~\ref{cross-sectional},
%\begin{equation*}
%\begin{split}
%&\sum_{j=1}^T \sum_{i=1}^n R_{ij}SW^{C\star}_{ij}({\blambda})\left\{(T-j+1)\widetilde{\bH}_{i, j-1}-(T-j)\text{pr}(C_{ij+1}=1\mid \overline{X}_{ij},C_{ij}=1;\hat{\eta})[1,\overline{X}_{ij}]\right\}\\
%=&(T-1)\sum_{i=1}^n\text{pr}(C_{i2}=1\mid X_{i1};\hat{\eta})[1,X_{i1}].
%\end{split}
%\end{equation*}

%Restrictions for the stabilized weights  can be derived, e.g.,  by multiplying the summands in~\eqref{censorbinary} by $\text{pr}(R_{i1}=1\mid A_{i0};\hat{\bet})$, 
%$$ \sum_{i=1}^n R_{i1}SW^{C\star}_{i1}({\blambda})\widetilde{\bH}_{i0}=\sum_{i=1}^n\text{pr}(R_{i1}=1\mid A_{i0};\hat{\bet})\widetilde{\bH}_{i0}, $$ where
%$SW^{C\star}_{i1}({\blambda})=W^C_{ij}(\hat{\btheta})\text{pr}(R_{i1}=1\mid A_{i0};\hat{\bet})c_{i1}(\blambda)$. 

\subsection{Related work}\label{hanswork}
In related work, \cite{Han2016} proposed to calibrate inverse probability of censoring weights by imposing similar restrictions to \eqref{missingrestrict}. However, the focus of \cite{Han2016} was on an eventual outcome at the end of study, $Y_{iT}$. For comparison, we  derive restrictions for his target of inference $\Ex(Y_{iT})$ as
\begin{equation}\label{missingeventual}
\sum_{j=1}^T \sum_{i=1}^n[R_{ij}W^{C\star}_{ij}({\blambda})-R_{i,j-1}W^{C\star}_{i,j-1}({\blambda})]\widetilde{\bH}_{i, j-1}=0, 
\end{equation}
which are based on~\eqref{missingliketimej} but with $T$ as the upper limit of the product in $k$. 
If $\widetilde{\bH}_{i, j-1}$ includes baseline covariates $\bV_i$, these restrictions impose 
$$\sum_{i=1}^nR_{iT}W^{C\star}_{iT}({\blambda})=n~~~~ \text{and} ~~~~ \sum_{i=1}^nR_{iT}W^{C\star}_{iT}({\blambda})\bV_i=\sum_{i=1}^n\bV_i,$$ 
 i.e., at visit $T$ the sample size after weighting is $n$ and the weighted average of $\bV_i$ is equal to the sample average of $\bV_i$. 
 
 The restrictions in~\eqref{missingeventual} and those in \cite{Han2016} differ in the way they achieve parsimony, which helps prevent unstable weights. In particular, \cite{Han2016} imposes separate restrictions at each visit, by multiplying the summands in~\eqref{missingeventual} by the visit indicator  $I(j=k)$ $(k=1,\ldots,T)$. This is feasible with few follow-up visits, e.g., 
 the simulation study in \cite{Han2016} has three visits.
Our approach leaves the degree of smoothing over time at the discretion of the researcher. For example, natural cubic splines can be incorporated to reflect non-linear time trend in the censoring process, similarly to the approach in   \cite{Hernan2001} for approximating the time-varying baseline treatment assignment distribution. Han also uses multiple estimates of $\Ex(Y_{iT}|\overline{H}_{i,j-1})$ based on different working models in place of $\widetilde{\bH}_{i, j-1}$ in \eqref{missingeventual}. Including more estimates of $\Ex(Y_{iT}|\overline{H}_{i,j-1})$ in the restrictions generally lead to an increase in efficiency when all working models are misspecified \citep{Han2016}. In this regard, Han's approach is more parsimonious than ours because his restrictions are contained in~\eqref{missingeventual} if his working models for $\Ex(Y_{iT}|\overline{H}_{i,j-1})$ are linear in $\widetilde{\bH}_{i, j-1}$ (see Section~\ref{Covariate} for more details).

\section{Joint calibrated estimation}
\subsection{Combining calibrated weights}\label{combiningrestrictions}
When both IPTW and IPCW are required for fitting MSMs, we propose to  perform calibration jointly and  the calibrated weights are  $W^{AC\star}_{ij}(\hat{\blambda})=SW^A_{ij}(\hat{\balpha},\hat{\bbeta})W^C_{ij}(\hat{\btheta})c(\overline{X}_{i,j-1},\overline{H}_{i,j-1},\hat{\blambda})$, where $c(\overline{X}_{i,j-1},\overline{H}_{i,j-1},{\blambda})$ is  a non-negative function satisfying $c(\overline{X}_{i,j-1},\overline{H}_{i,j-1},{\blambda}=\mathbf{0})=1$, and $\hat{\blambda}$ is the estimate of ${\blambda}$ such that $W^{AC\star}_{ij}(\hat{\blambda})$ satisfies the restrictions~\eqref{treatrestrict} and~\eqref{missingrestrict2}. Since~\eqref{missingrestrict2} already imposes constraints on the  sample size at each visit after weighting,~\eqref{normalizerestrict} is excluded to avoid collinearity. 

Another approach is to use the product of the individually calibrated weights $W^{AC\star}_{ij}(\hat{\blambda})=SW^{A\star}_{ij}(\hat{\blambda})W^{C\star}_{ij}(\hat{\blambda})$. However, while $SW^{A\star}_{ij}(\hat{\blambda})$ and $W^{C\star}_{ij}(\hat{\blambda})$ satisfy the restrictions~\eqref{treatrestrict}--\eqref{normalizerestrict} and~\eqref{missingrestrict2} respectively, their observation-specific product may not. Thus these estimated weights may lose the potential benefits from calibration.  

\subsection{Implementation of the calibration procedure}\label{Implementation}
%Implementation of the joint calibration requires specifying a non-negative function $c_{ij}(\blambda)$ and a method for solving  the restrictions to obtain $\hat{\blambda}$. We propose to use an exponential function for $c_{ij}(\blambda)$ because it can exploit the fact that the restrictions~\eqref{treatrestrict}--\eqref{normalizerestrict} and~\eqref{missingrestrict2} are linear in the calibrated weights. Specifically

We collect all initial and calibrated weights into $m\times 1$ vectors $W(\hat{\balpha},\hat{\bbeta},\hat{\btheta})$ and $W^{\star}(\blambda)$, respectively, where  $m$ is the number of weights. If no censoring occurs, then  $m=nT$. The implementation of the joint calibration requires solving a system of linear equations in terms of $W^{\star}(\blambda)$ since the restrictions~\eqref{treatrestrict}--\eqref{normalizerestrict} and~\eqref{missingrestrict2} are linear in the calibrated weights. Let $\mathit{K}$ be the \emph{known} $m\times r$ matrix and ${\mathbf{l}}$ be the \emph{known}  $r \times 1$ vector, where $r$ is the numbers of restrictions. For example,  for IPTCW $r$ would be the combined size of $\bbeta_\text{w}$ and $\btheta_\text{w}$.  Both $\mathit{K}$ and ${\mathbf{l}}$ are  determined by the calibration restrictions.
Broadly, for obtaining the calibrated weights,  we would  like to solve
\begin{equation}\label{restrictionexample}
\mathit{K}^\top W^{\star}(\blambda)- \mathbf{l}=0.
\end{equation}
We propose the calibration of the form  $W^{\star}(\blambda)=W(\hat{\balpha},\hat{\bbeta},\hat{\btheta}) \circ \exp(K\blambda)$, where $\exp(\cdot)$ is performed element-wise, $\circ$ denotes element-wise product, and $\blambda$ is a $r\times 1$ vector of parameters to be estimated. The choice of this calibration function is motivated by the equivalence between solving~\eqref{restrictionexample} and minimizing the convex function for $\blambda$,
\begin{equation}\label{objectiveexample}
\mathbf{1}^\top \{W(\hat{\balpha},\hat{\bbeta},\hat{\btheta})\circ \exp({\mathit{K}}\blambda)\}-\mathbf{l}^\top\blambda,
\end{equation}
where $\mathbf{1}$ is an $m\times 1$ vector of ones. The convexity of~\eqref{objectiveexample} ensures that the solution to~\eqref{restrictionexample} is unique and can be found efficiently, particularly when the $r\times r$ Hessian matrix is used in the estimation of $\blambda$.  Specifically, the $j$th column of the Hessian matrix is $\mathit{K}^\top\{K_{\cdot j}\circ W(\hat{\balpha},\hat{\bbeta},\hat{\btheta})\circ \exp(\mathit{K}\blambda)\}$, where $K_{\cdot j}$ is the $j$th column of $\mathit{K}$.
We minimize~\eqref{objectiveexample} by solving~\eqref{restrictionexample} using the R \citep{RCoreTeam2014} package \texttt{nleqslv} \citep{Hasselman2016}.

\subsection{Choice of models for deriving restrictions}\label{Covariate}
In this section, we provide a discussion on how to choose the  models $\text{pr}(A_{ij}\mid \overline{X}_{i,j-1};\bbeta_{\text{w}})$ and $\text{pr}(R_{ij}=1\mid \overline{H}_{i,j-1},R_{i,j-1}=1;\btheta_\text{w})$ to derive restrictions~\eqref{treatrestrict} and~\eqref{generalmissingrestrict}.
Here we emphasize again that the models for deriving the restrictions and the models for estimating the initial weights can be different. 

 First, as long as
 \bnu
 \item[(a)] the models for deriving restrictions are parameterized such that $\text{pr}(A_{ij}\mid \overline{X}_{i,j-1};\bbeta_{\text{wb}}=\hat{\balpha},\bbeta_{\text{wd}}=\mathbf{0})=\text{pr}(A_{ij}\mid \overline{A}_{i,j-1};\hat{\balpha})$  (here $\text{pr}(A_{ij}\mid \overline{A}_{i,j-1};\hat{\balpha})$ are the terms in the numerator of the initial stabilized treatment weights) and $\text{pr}(R_{ij}=1\mid \overline{H}_{i,j-1},R_{i,j-1}=1;\btheta_\text{w}=\mathbf{0})=1/2$, respectively,
 \item[(b)] the models for the initial weights $\text{pr}(A_{ij}\mid \overline{X}_{i,j-1};\bbeta)$ and $\text{pr}(R_{ij}=1\mid \overline{H}_{i,j-1},R_{i,j-1}=1;\btheta)$ have been correctly specified,
 \enu  the choice of parametrizations for $\text{pr}(A_{ij}\mid \overline{X}_{i,j-1};\bbeta_{\text{w}})$ and $\text{pr}(R_{ij}=1\mid \overline{H}_{i,j-1},R_{i,j-1}=1;\btheta_\text{w})$ does not affect the consistency of the IPWEs. 
This is because the initial weights will  converge to the true weights, and they themselves satisfy the population versions of the restrictions in~\eqref{treatrestrict}--\eqref{normalizerestrict} and~\eqref{generalmissingrestrict} (see the Supplementary Material for proof). Thus $\hat{\blambda}$ will converge to $\mathbf{0}$, i.e., no calibration is  applied, thereby maintaining the consistency of the IPWEs with the initial weights. We recommend choosing treatment process models for deriving restrictions that can take the values of the numerator of the initial stabilized weights, while logistic models, i.e., the restrictions~\eqref{missingrestrict}, will suffice for the censoring process.

Second, we distinguish between  covariate histories that are predictive of $\Ex(Y^{\overline{a}_{j},\overline{r}_{j}=\mathbf{1}}_{ij})$, denoted as $\overline{X}^Y_{i,j-1}$, and those that are predictive of $A_{ij}$, denoted as $\overline{X}^A_{i,j-1}$. Some elements of $\overline{X}^Y_{i,j-1}$ and $\overline{X}^{A}_{i,j-1}$ overlap which leads to confounding bias, while others may be distinct from each other. We recommend prioritizing $\widetilde{\bX}^Y_{i,j-1}$, i.e., functionals of $\overline{X}^Y_{i,j-1}$, for inclusion in the models of the treatment and censoring processes for deriving restrictions at visit $j$. This is because, conditional on $\overline{A}_{i,j-1}$, it is precisely $\overline{X}^Y_{i,j-1}$ that induces confounding bias, as they confound the effect of $a_{j}$ on $\Ex(Y^{{a}_{j},{r}_{j}=1}_{ij} \mid \overline{A}_{i,j-1}, R_{i,j-1}=1)$.  $\overline{X}^Y_{i,j-1}$ also induces  selection bias if their distribution changes when conditioning on being uncensored. 

Our recommendation is partly supported by the fact that the left-hand side of~\eqref{missingeventual} is equivalent to the augmentation term in the augmented inverse probability weighted estimator (AIPWE) for estimating $\Ex(Y_{iT})$ in the presence of censoring \citep{Robins1995a,Robins1995b,Rotnitzky1995} if $\widetilde{\bH}_{i,j-1}$ is replaced by an estimate of $\Ex(Y_{iT}\mid \overline{X}^Y_{i,j-1})$. In particular, if the estimate of $\Ex(Y_{iT} | \overline{X}^Y_{i,j-1})$ is linear in $\widetilde{\bX}^Y_{i,j-1}$, and $\widetilde{\bH}_{i,j-1}$  contains $\widetilde{\bX}^Y_{i,j-1}$, then our estimator based on the calibrated weights can exploit the same information in the observed data as the AIPWE. This is because the augmentation term based on our calibrated weights will be \emph{zero }by definition of the restrictions. Furthermore, for estimating the average treatment effect in the cross-sectional setting,  \cite{Zhao2016} find that balancing all pre-treatment covariates in the outcome model is more efficient than balancing only pre-treatment covariates in the treatment model  when constructing covariate balancing weights for IPTW.
%find that balancing all pretreatment covariates in the outcome model $\Ex(Y^{a}_i\mid \bX^Y_i)$, $\widetilde{\bX}^Y_i$, across treatment levels is more efficient than balancing only pretreatment covariates in the treatment model $\text{pr}(A_{i}=1\mid {\bX}^A_i)$, $\widetilde{\bX}^A_i$, when constructing covariate balancing weights for IPTW.

\section{Simulation study}
We conduct a simulation study to assess the finite sample performance of the IPWE for MSMs based on our calibrated estimation approach and  the maximum likelihood approach for weight estimation. 

\subsection{Design}
The design of the simulation studies is motivated by  the HERS data, where the time-varying treatment is an ordinal variable. The data generating mechanism for a patient is summarized in Table~\ref{simset} and  Figure \ref{fig_sim} provides a pictorial description. We omit the subscript $i$ for patients for clearer presentation. 
\begin{table}[h!]
\centering
\caption{Data generating mechanism for the simulations\label{simset}} 
\begin{tabular}{llll}
\hline
\textbf{Baseline} ($j=0$)&          &                                  &\\
 \multicolumn{1}{r}{\small \textit{Censoring}:~} & $R_0=1$   &&\\
 \multicolumn{1}{r}{\small \textit{Treatment}:~} & $A^0_0 \sim \text{Bernoulli}(0.5)$, &&\\ 
                                 & $A^1_0\mid A^0_0=1\sim \text{Bernoulli}(0.5)$ & \\
    \multicolumn{1}{r}{\small \textit{Covariates}:} & $X_{01}=U_0Z_{01}$,    & & \\
                                    & $X_{02}=U_0Z_{02}$,     & &\\                              
                                     &  $X_{03}=Z_{03}+0.5(A^0_0+A^1_0)$, &&\\
                                     & $X_{04}=Z_{04}+0.5(A^1_0+A^1_0)$, &&\\
                                     &  where $U_0=1-0.3(A^0_0+A^1_0)$, &  \multicolumn{2}{l}{ $Z_{01},Z_{02},Z_{03},Z_{04} \stackrel{i.i.d}{\sim} N(0,1)$} \\
                                     &&&\\
  \textbf{Follow-up visits} & ($j=1,\ldots,10$)&                                          &\\
  \multicolumn{1}{r}{\small \textit{Censoring}:~} & $R_j \mid R_{j-1}=1  \sim \text{Bernoulli}(q_j)$   &&\\
                                   & \multicolumn{3}{l}{{\small \emph{Scenario (1)}}: $q_j=1$ (no censoring)} \\
                                  & \multicolumn{3}{l}{{\small \emph{Scenario (2)}}: $\logit(q_j)= 1+A^0_{j-1}+A^1_{j-1}+0.5X_{j-1,1}+0.5X_{j-1,2}$} \\
                                  & \multicolumn{3}{l}{ $~~~~~~~~~~~~~~~~~~~~~~~~~~~~~~~~+0.2X_{j-1,3}+0.2X_{j-1,4}$} \\
   \multicolumn{1}{r}{\small \textit{Treatment}:~} & $A^0_j \sim \text{Bernoulli}(p_j)$, &&\\ 
                                 & $A^1_j\mid A^0_j=1\sim \text{Bernoulli}(p_j)$ & \\                                
    & \multicolumn{3}{l}{$\logit(p_j)=A^0_{j-1}+A^1_{j-1}+0.5X_{j-1,1}+0.5X_{j-1,2}-0.2X_{j-1,3}-0.2X_{j-1,4}$} \\
    \multicolumn{1}{r}{\small \textit{Covariates}:} & $X_{j1}=U_jZ_{j1}$,    & & \\
     & $X_{j2}=U_{j}Z_{j2}$,     & &\\                              
                                     &  $X_{j3}=Z_{j3}+0.5\sum_{t=0}^{j}(A^0_t+A^1_t)$, &&\\
                                     & $X_{j4}=Z_{j4}+0.5\sum_{t=0}^j(A^0_t+A^1_t)$, &&\\
                                     &  where $U_{j}=1-0.3(A^0_j+A^1_j)$, &\multicolumn{2}{l}{ $Z_{j1},Z_{j2},Z_{j3},Z_{j4} \stackrel{i.i.d}{\sim} N(0,1)$}\\
                                      \multicolumn{1}{r}{\small \textit{Outcome}:~~~}& \multicolumn{2}{l}{$Y_j=200+5(A^0_j+A^1_j+\sum_{t=j-1}^j\sum_{l=1}^4 X_{tl})+\epsilon_{j},~\epsilon_{j}\sim N(0,20)$}\\
      \hline
\end{tabular}
\end{table}

\begin{figure}[h!]

\centering
\begin{tikzpicture}
\tikzstyle{connect}=[->,shorten >=1pt,node distance=10cm,semithick]
  \node (0)  {$A_1$};
\node (1) [right=1cm of 0] {$A_{T-1}$};
  \node (2) [right=1cm of 1] {$X_{T-1}$};
  \node (3) [right=1cm of 2] {$A_T$};
\node (4) [below right=0.75cm of 2] {$R_{T}$};
%\node (5) [below=1cm of 2] {$Y_{T-1}$};
\node (6) [right=1cm of 3] {$X_T$};
\node (7) [below right=0.75cm of 6] {$Y_{T}$};

\path (0) -- node[auto=false]{$\ldots$}(1);
\path (0) edge [connect, bend left] (2);
\path (0) edge [connect, bend left] (6);
\path (1) edge [connect] (2);
\path (1) edge [connect] (4);
\path (1) edge [connect, bend left] (3);
\path (1) edge [connect, bend left] (6);
%\path (1) edge [connect] (5);
\path (2) edge [connect] (3);
\path (2) edge [connect] (4);
%\path (2) edge [connect] (5);
\path (2) edge [connect] (7);
\path (3) edge [connect] (6);
\path (3) edge [connect] (7);
\path (6) edge [connect] (7);

\end{tikzpicture}
\caption{
The relationship between the variables in the simulation set-up.}
\label{fig_sim}
\end{figure}
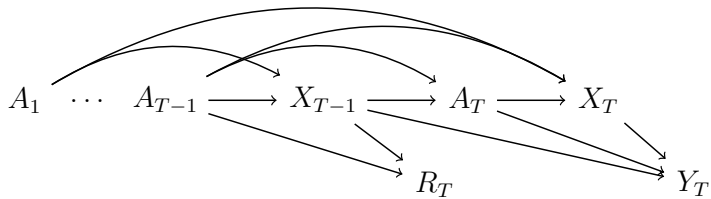

%\begin{equation*}
%\begin{split}
%&R_0=1, , .\\
%&X^1_1=U_1Z^1_1,X^2_1=U_1Z^2_1,X^3_1=Z^3_1+0.5(A^0_1+A^1_1),X^4_1=Z^4_1+0.5(A^1_0+A^1_1),\\
%\end{split}
%\end{equation*}
%where $U_1=1-0.3(A^0_1+A^1_1)$ and $Z^1_1,Z^2_1,Z^3_1,Z^4_1\sim N(0,1)$. Then for $j=2,\ldots,11$
%\begin{equation*}
%\begin{split}
%&C_j\sim \text{Bernoulli}(q_j),~\text{where for scenario 1}~q_j=1,~\text{and for scenario 2}\\
%&q_j=1-\text{expit}(-1-A^0_{j-1}-A^1_{j-1}-0.5X^1_{j-1}-0.5X^2_{j-1}-0.2X^3_{j-1}-0.2X^4_{j-1}).\\
%&A^0_j\sim \text{Bernoulli}(p_j),A^1_j\mid A^0_j=1\sim \text{Bernoulli}(p_j),\\
%&p_j=\text{expit}(A^0_{j-1}+A^1_{j-1}+0.5X^1_{j-1}+0.5X^2_{j-1}-0.2X^3_{j-1}-0.2X^4_{j-1}).\\
%&X^1_{j}=U_jZ^1_{j},X^2_{j}=U_{j}Z^2_{j},X^3_{j}=Z^3_{j}+0.5\sum_{k=0}^{j}(A^0_k+A^1_k),X^4_{j}=Z^4_{j}+0.5\sum_{k=0}^j(A^0_k+A^1_k).\\
%&Y_j=200+5(A^0_j+A^1_j+\sum_{k=j-1}^j\sum_{l=1}^4 X^l_k)+\epsilon_{j},~\epsilon_{j}\sim N(0,20),
%\end{split}
%\end{equation*}
%where $U_{j}=1-0.3(A^0_j+A^1_j)$ and $Z^1_{j},Z^2_{j},Z^3_{j},Z^4_{j}\sim N(0,1)$.

In this set-up there are four time-varying confounders $\{X_{j-1,1},X_{j-1,2},X_{j-1,3},X_{j-1,4}\}$ $(j=1,\ldots,10)$, which   affect the treatment assignment $\{A^0_j,A^1_j\}$ and  the mean of the outcome $\Ex(Y_j)$ at visit $j$. 
In addition, the variances of $\{X_{j-1,1},X_{j-1,2}\}$ and  the means of $\{X_{j-1,3},X_{j-1,4}\}$ are affected by  treatment history $\overline{A}_{j-1}$. 
Time-varying confounding   arises from $\{X_{j-1,3},X_{j-1,4}\}$ because conditioning on them via regression adjustment blocks the effect of previous treatments  (i.e.,  $\sum_{t=0}^{j-1}(A^0_t+A^1_t)$) through themselves. 

For the censoring process, we specify two scenarios. In Scenario (1), no censoring occurs as $\text{pr}(R_j \mid R_{j-1}=1)=1$. In Scenario (2), covariate-dependent censoring occurs 
and selection bias is induced because $\Ex(Y_j)$ depends on $X_{j-1,l}$ and $\Ex(X_{j-1,l}\mid \overline{A}_{j-1},R_{j}=1)\neq \Ex(X_{j-1,l}|\overline{A}_{j-1})$ $(l=1,\ldots,4)$.

We  assume a MSM $\Ex(Y^{\overline{a}_{j}}_{j})=\gamma_0+\gamma_1\sum_{t=0}^j(a^0_{t}-a^1_{t})+\gamma_2\sum_{t=0}^ja^1_{t}$, where $a^0_{t}$ and $a^1_{t}$ are potential values of treatment indicators $A^0_{t}$ and $A^1_{t}$. The true treatment effects can be derived by noting that $\Ex(Y_j|\overline{A}_j)=200+10\sum_{t=0}^j(A^0_t+A^1_t)$, and thus $\gamma_1=10$  and $\gamma_2=20$.

Data from each patient are generated independently. We simulate 2500 data sets with 500, 1000 and 2500 patients and  10 scheduled follow-up visits after baseline. For weight estimation, we assume the logistic models~\eqref{ordinaltreatment} for the treatment indicators and the logistic model for the censoring process as in Section~\ref{censorcal}. We apply both the maximum likelihood approach and the proposed calibrated estimation approach by imposing the restrictions in~\eqref{normalizerestrict},~\eqref{ordinaltreatmentrestric} and~\eqref{missingrestrict2} on the initial weights obtained by maximum likelihood. Then we apply IPTW for Scenario (1) and IPTCW for Scenario (2) with the estimated weights and use the estimating equations in~\eqref{MSM} to estimate $\gamma_1$ and $\gamma_2$. 

In Scenario 1, we include an intercept and the main effects of $\{A^0_{j-1},A^1_{j-1}\}$ in the logistic models for the numerator of the stabilized treatment weights at visit $j$. For the denominator, we additionally include the main effects of $\{X_{j-1,1},X_{j-1,2},X_{j-1,3},X_{j-1,4}\}$ to ensure that the treatment assignment models are correctly specified. In Scenario 2, we use the same treatment assignment models as Scenario 1, and include visit-specific indicators and the main effects of $\{A^0_{j-1},A^1_{j-1},X_{j-1,1},X_{j-1,2},X_{j-1,3},X_{j-1,4}\}$ in the censoring model at visit $j$. The visit-specific intercepts were included to impose analogous restrictions to~\eqref{normalizerestrict} in Scenario 2. To consider a functional form misspecification, we use a set of transformed covariates  $\{X^{t}_{j-1,1},X^{t}_{j-1,2},X^{t}_{j-1,3},X^{t}_{j-1,4}\}$ of the form $X^{t}_{j-1,1}=(X_{j-1,1})^3/9$, $X^{t}_{j-1,2}=X_{j-1,1}X_{j-1,2}$, $X^{t}_{j-1,3}=\log(|X_{j-1,3}|)+4$ and $X^{t}_{j-1,4}=\exp(X_{j-1,4})/\{1+\exp(X_{j-1,4})\}$ in place of the correct covariates $\{X_{j-1,1},X_{j-1,2},X_{j-1,3},X_{j-1,4}\}$ for estimating the initial weights and for deriving restrictions for calibration.

%We do this separately for the case where the correct covariates, $\{X^1,X^2,X^3,X^4\}$, and transformed covariates $\{X^{1t},X^{2t},X^{3t},X^{4t}\}$ of the form $X^{1t}=(X^1)^3/9$, $X^{2t}=X^1X^2$, $X^{3t}=\log(|X^3|)+4$ and $X^{4t}=\text{expit}(X^4)$ are used to construct the base weights and restrictions. 

%As an aside, the causal effect in this set-up can be more efficiently estimated by IPTW with weights $\text{pr}(A_{ij}\mid \overline{A}_{i,j-1})/\text{pr}(A_{ij}\mid \overline{X}_{i,j-1})$, which are sufficient to account for confounding because $A_{ij}$ is the only treatment variable in $E(Y_{ij}\mid \overline{X}_{ij})$ that has its effect on $E(Y_{ij})$ confounded. The trade-off is a lost of consistency when other treatment effects are confounded in $E(Y_{ij}\mid \overline{X}_{ij})$. For the calibration approach, the restrictions for these alternative weights are those in~\eqref{treatrestrict} but with $j$ replacing the lower limit of the summation in $k$.

\subsection{Results}
Table~\ref{Simtable} summarizes the results of the simulation studies.
When the models for the treatment and  censoring processes are correctly specified, i.e., when the correct covariates are used, it is not surprising that  
the biases from IPWEs  with  weights based on maximum likelihood and with calibrated weights are negligible and the standard deviations decrease as  sample size increases. However, the IPWEs  with calibrated weights have smaller standard deviations and mean squared errors than the IPWEs with weights from maximum likelihood because the calibration reduces the finite-sample estimation error by optimising covariate balance.

 \begin{table}[h!]
\centering
\caption{Empirical bias, standard deviation (SD) and root mean squared error (RMSE) of the parameter estimates in the marginal structural model  from applying inverse probability of treatment weighting  and inverse probability of treatment and censoring weighting to Scenarios (1) and (2), respectively,  in the simulation study.   The weights are based on the maximum likelihood (MLE) and  calibration (CMLE) approaches.}
\label{Simtable}
\small
\begin{tabular}{ccccccc}
\hline
\hline
&\multicolumn{3}{c}{Scenario 1}&\multicolumn{3}{c}{Scenario 2}\\

 &Bias &SD &RMSE &Bias &SD &RMSE \\
&  ($\gamma_1$, $\gamma_2$) & ($\gamma_1$, $\gamma_2$)& ($\gamma_1$, $\gamma_2$)& ($\gamma_1$, $\gamma_2$) & ($\gamma_1$, $\gamma_2$)& ($\gamma_1$, $\gamma_2$)\\
\hline
\multicolumn{3}{l}{$n=500$} & & & & \\
\multicolumn{7}{c}{\textit{correct covariates}}\\
MLE &$-$0.00~~0.02&0.71~~0.67&0.71~~0.67&$-$0.02~~0.00&1.19~~1.01&1.19~~1.01\\
CMLE &$-$0.01~~0.00&0.63~~0.59&0.63~~0.59&$-$0.03~~$-$0.01&1.12~~0.95&1.12~~0.95\\
\multicolumn{7}{c}{\textit{transformed covariates}} \\
MLE  &0.19~~0.33&1.67~~1.15&1.68~~1.19&0.10~~0.25&2.15~~1.44&2.15~~1.46\\
CMLE &0.12~~0.29&0.52~~0.43&0.53~~0.52&0.07~~0.17&0.94~~0.77&0.95~~0.79\\
 & & & & & &\\

\multicolumn{3}{l}{$n=1000$} & & & & \\
\multicolumn{7}{c}{\textit{correct covariates}}\\
MLE &0.01~~0.01&0.51~~0.56&0.51~~0.56&$-$0.02~~0.00&0.99~~0.81&0.99~~0.81\\
CMLE &0.01~~0.02&0.48~~0.47&0.48~~0.47&$-$0.02~~$-$0.01&0.85~~0.72&0.85~~0.72\\

\multicolumn{7}{c}{\textit{transformed covariates}} \\
MLE  &0.13~~0.29&1.59~~1.46&1.60~~1.49&0.05~~0.30&1.83~~1.71&1.83~~1.73\\
CMLE &0.13~~0.27&0.41~~0.34&0.43~~0.44&0.09~~0.17&0.75~~0.61&0.76~~0.63\\
 & & & & & &\\
\multicolumn{3}{l}{$n=2500$} & & & & \\
\multicolumn{7}{c}{\textit{correct covariates}}\\

MLE &$-$0.00~~$-$0.01&0.34~~0.40&0.34~~0.40&0.02~~0.00&0.69~~0.63&0.69~~0.63\\
CMLE &$-$0.00~~$-$0.01&0.31~~0.33&0.31~~0.33&0.01~~$-$0.00&0.59~~0.51&0.59~~0.51\\

\multicolumn{7}{c}{\textit{transformed covariates}} \\
MLE  &0.20~~0.28&2.20~~2.09&2.21~~2.11&0.17~~0.33&2.62~~2.62&2.62~~2.65\\
CMLE &0.14~~0.26&0.31~~0.27&0.34~~0.38&0.07~~0.15&0.57~~0.47&0.57~~0.49\\
\hline
\end{tabular}
\end{table}

 In contrast, when the models for the treatment and  censoring processes are misspecified, i.e., when the transformed covariates are used, the IPWEs with  weights from maximum likelihood and with calibrated weights both  have  non-negligible biases that do not decrease with increasing sample size, although for most of the scenarios, the IPWEs with calibrated weights have slightly smaller biases. However, the IPWEs with calibrated weights are more efficient with much lower standard deviations. As a result,  the IPWEs with calibrated weights have much smaller mean squared errors than the IPWEs with weights from maximum likelihood. 
 
A more alarming feature of the IPWEs with weights from maximum likelihood is that the large standard deviations and  mean squared errors  even increase with sample size. Under model misspecification, this occurred because a few sets of estimated weights from maximum likelihood exacerbated the extremeness of the tails of the sampling distribution of the parameter estimators as sample size increased. \citet[pp. 553--4]{Robins2007} give more details of this phenomenon.
In contrast, the IPWEs with calibrated weights do not exhibit this undesirable property and  show  more robustness to the functional form misspecification  in our simulation set-up. 
 
 Overall, the simulation results show that the proposed calibration approach  can improve the  efficiency of the IPWEs for MSMs and provide more robustness to functional form misspecification.
 
\subsection{Theoretical explanations}\label{theory}
The above empirical findings about the performance of the IPWEs with calibrated weights can be explained by the recent theoretical results in  \cite{Tan2017}. \cite{Tan2017} shows that even  under model misspecification, calibration with restrictions on covariate balance can reduce the relative error of the estimated weight compared to the true weights, which  controls  the mean squared error of the IPWE. The maximum likelihood approach for weight estimation focuses on reducing the absolute error of the estimated weights compared to the true weights, which is not directly connected to the mean squared error of the IPWE. Therefore, under model misspecification calibrating the weights in terms of covariate balance can still reduce the mean squared error of the IPWE and improve its  efficiency. However, since bias is quantified by averaging over repeated samples, depending on specific set-up for model misspecification, the IPWE with weights from maximum likelihood can have similar or smaller bias than the IPWE with calibrated weights because large positive and negative differences from the true parameter values can be cancelled out when averaging across samples.

 In the Supplementary Material, we also show why the maximum likelihood approach may perform poorly in terms of 
 removing covariate imbalances asymptotically  
 under mild model misspecification. Consequently, these covariate imbalances lead to poor performance of the corresponding IPWE even with large sample size.

\section{Application to the HERS data}\label{hersexample}

\subsection{Description of the  cohort}

At the early period of the HERS from 1993 to 1995, treatment with a single ART  was recommended for patients with CD4 cell counts less than 500. Beginning in late 1995 and early 1996, HAART (a combination of three or more ARTs) became more widely used  in the HERS cohort. This increasing use of HAART is reflected in the right panel of  Figure~\ref{Fig1}, which presents the crude percentages of patients who received HAART over the follow-up visits.  Since the enrolment period of the HERS was between 1993 and 1995, the upward trend of HAART use started at visit 5, roughly two and half years into the study. The left panel of  Figure~\ref{Fig1} also presents the sample averages of the CD4 counts at each visit, which  show a decreasing trend up to  visit 5. Then there is a level-off which coincides with the widespread use of HAART in this cohort. This phenomenon does not necessarily suggest the efficacy of HAART because it could be due to selection bias from dropout of patients with severe disease progression or other factors.

\begin{figure}[h!]
\includegraphics[scale=0.85]{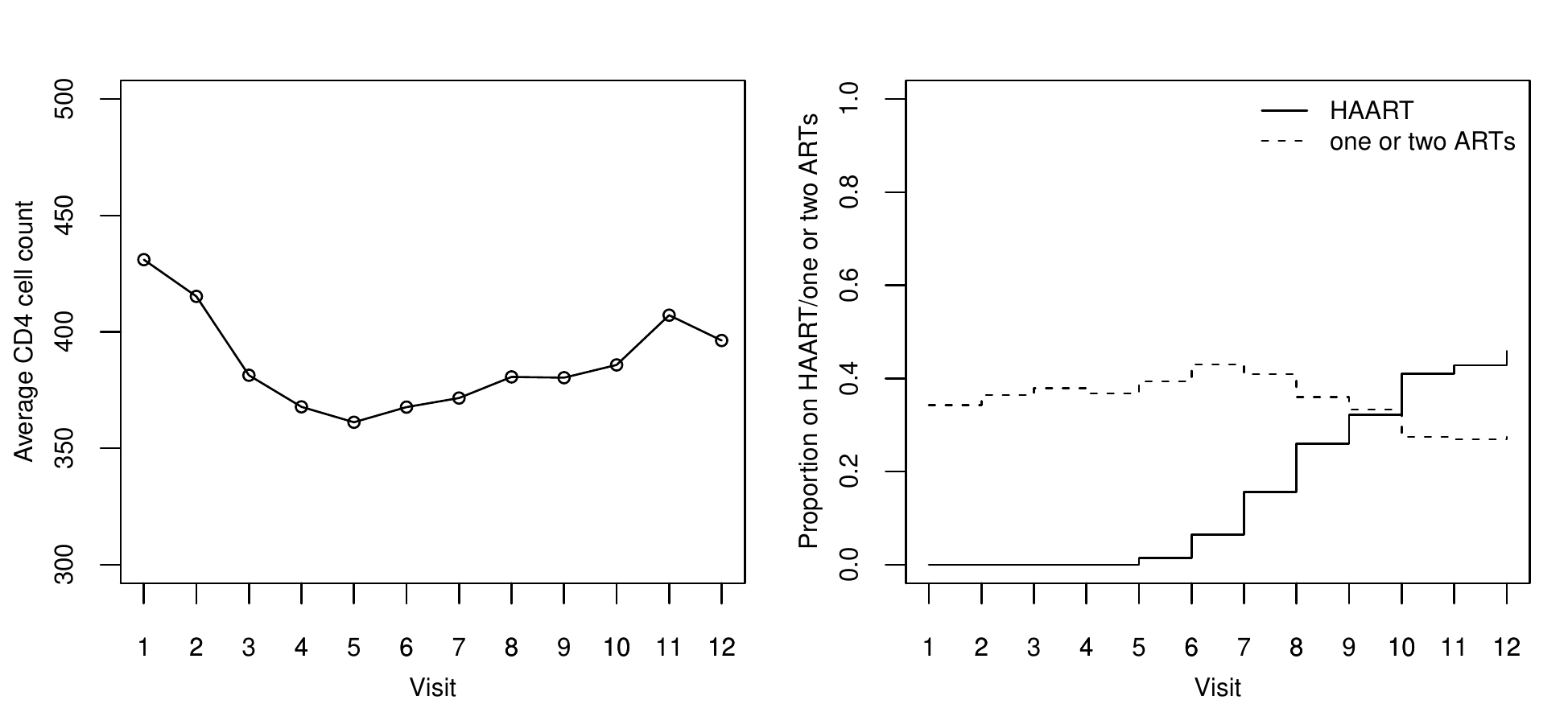}
\centering
\caption{The average CD4 cell count and the proportion of patients who received  HAART and one or two ARTs at each visit in the HERS data.}
\label{Fig1}
\end{figure}

%\begin{table}[h!]
%\centering
%\caption{The number of patients who were HIV-positive at enrolment and still under follow-up  at each visit in the HERS data.}
%\label{Freqtable}
%\small
%\begin{tabular}{lcccccccccccc}
%\hline
%\hline
%Visit&1&2&3&4&5&6&7&8&9&10&11&12\\
%$\#$ of patients &843&779&750&717&668&636&610&584&563&532&507&395\\
%\hline
%\end{tabular}
%\end{table}

\begin{table}[h!]
\centering
\caption{The number of patients who were HIV-positive at enrolment and still under follow-up, had their CD4 count observed, and dropped out at each visit in the HERS data.}
\label{Freqtable}
\small
\begin{tabular}{lcccccccccccc}
\hline
\hline
Visit&1&2&3&4&5&6&7&8&9&10&11&12\\
\hline
In follow-up &871&815&787&748&702&668&636&614&595&571&553&449\\
CD4 observed &850&706&692&665&617&587&576&547&522&506&492&405\\
Dropped out &56&28&39&46&34&32&22&19&24&18&104&133\\
\hline
\end{tabular}
\end{table}

As mentioned, attrition by dropout in the HERS is substantial. Table~\ref{Freqtable} displays the number of patients who were HIV-positive at enrolment and still under follow-up, had their CD4 count observed, and dropped out at each visit. Besides attrition, there were secondary sources of missing data which resulted in intermittent missing data (before being lost to follow-up), missing data at enrolment for CD4 counts, and left-censored HIV viral load at the lower detection limit (LDL).  We deal with these by following the approaches in \cite{Ko2003}. Specifically, if there was one intermittent missing CD4 count value then the last observed value was carried forward; otherwise the patient was treated as having dropped out between their last observed visit and the next visit. Missing CD4 values at enrolment were imputed from the patient's second visit, if possible; otherwise the patients missing CD4 values at visits 1 and 2 were excluded from the analysis. Finally, left-censored viral load values were imputed from a uniform distribution on the interval $[0,\text{LDL}]$, where $\text{LDL}$ was set at either 50 ($13\%$ of the viral load observations) or 500 ($1\%$ of the viral load observations) depending on the assays used.

\subsection{Model parameterizations and estimation}
%In a previous analysis, \cite{Ko2003} estimated the effect of HAART on CD4 count from visits 8 to 12 with similar data. In their analysis, they defined women not on HAART as those who had one or two ARTs or no treatment. Therefore their estimate may have been an underestimate of the therapeutic effect of HAART relative to no treatment. 

%In contrast, we make the distinction in treatments for those not on HAART in our marginal structural model. 

Since HAART was not available at enrolment in the HERS cohort, we follow \cite{Ko2003} and treat visit 7, when HAART was more widely used in the HERS, as the  `baseline' and estimate the causal effects of HAART over the two-year period between visit 8 and visit 12. In total, there are 610 patients at visit 7 who had at least one CD4 count measured between visit 8 and 12 and sufficient information for covariates to estimate the weights for IPTW and IPTCW.  The total number of CD4 count observations for analysis is 2581.

As discussed in Section~\ref{exampleintro}, in order to provide a more precise estimate of the causal effect of HAART relative to no treatment,  we treat the time-varying antiviral  treatment as an ordinal variable  with 3 levels: `no treatment', `ART other than HAART' and `HAART'. Furthermore, we stratify the treatment effects within categories of CD4 count at visit 7, which are coded as $D_i=0$ if $Y_{i7}<200$, $D_i=1$ if $200\leq Y_{i7}\leq 500$ and $D_i=2$ if $Y_{i7}>500$, where $Y_{i7}$ is the CD4 count at visit 7. 

Specifically, we assume the following MSM
\begin{equation*}
\Ex(Y^{\overline{a}_{j}}_{ij})=\delta_{0j}+\sum_{k=1}^2\delta_k I(D_i=k)+\bdelta^\top_{v}\bV_{i}+\sum_{k=0}^2I(D_i=k)\left\{\gamma_{1k}\sum_{l=8}^j(a^0_{l}-a^1_{l})+\gamma_{2k}\sum_{l=8}^ja^1_{l}\right\}
\end{equation*}
for $j=8,\ldots,12$, where $\delta_{0j}$ are visit-specific intercept terms, $\bV_{i}$ are baseline covariates  evaluated at visit 7, and $\bdelta_v$ are their corresponding regression coefficients. For $\bV_{i}$, we include the following variables at visit 7: $\log_{10} $ HIV viral load, HIV symptom level (5-point scale), status of one or two ARTs and status of HAART. We also estimate an overall treatment effect by constraining $\gamma_{1k}$ and $\gamma_{2k}$ to be constant across $k$ for the baseline CD4 count level $D_i$. An additional MSM for evaluating the short-term treatment effect is also considered and presented in the Supplementary Material.

%Similarly to \cite{Ko2003}, we specify our marginal structural model from visits 8 to 12 and treat visit 7 as baseline, corresponding to the first visit where HAART was widely available. Furthermore, we stratify our treatment effects within categories of CD4 count at visit 7 defined by $D_i=0$ if $Y_{i7}<200$, $D_i=1$ if $200\leq Y_{i7}\leq 500$ and $D_i=2$ if $Y_{i7}>500$, where $Y_{i7}$ is the CD4 count at visit 7. Overall, our marginal structural model is
%\begin{equation*}
%E(Y^{\overline{a}_{ij}}_{ij})=\delta_{0j}+\sum_{k=1}^2\delta_k\mathbf{1}(D_i=k)+\delta^\top_{v}V_{i7}+\sum_{k=0}^2\mathbf{1}(D_i=k)\left\{\gamma_{1k}\sum_{l=7}^j(a^0_{il}-a^1_{il})+\gamma_{2k}\sum_{l=7}^ja^1_{il}\right\}
%\end{equation*}
%for $j=8,\ldots,12$, where $\delta_{0j}$ are time-specific intercept parameters, $V_{i7}$ are baseline covariates, and $\delta_v$ are their corresponding regression coefficients. For $V_{i7}$, we include the following variables at visit 7; $\log_{10}(\text{Viral load})$ (measurement of quantity of HIV), HIV symptom scale (5-point scale), one or two ARTs status and HAART status. We also estimate an overall treatment effect by constraining $\gamma_{1k}$ and $\gamma_{2k}$ to be independent of $k$. We consider an additional marginal structural model in the Supplementary Material.

The parameters in the MSMs were estimated by applying IPTW and IPTCW, with weights estimated by maximum likelihood and by applying the proposed  calibration. For IPTW, the treatment  model~\eqref{ordinaltreatment} was used to obtain maximum likelihood estimates of the weights and to derive restrictions~\eqref{ordinaltreatmentrestric} for calibration. 
%in $\widetilde{\bX}_{ij}^0$ and $\widetilde{\bX}_{ij}^1$ 
For the numerator in the SIPTW, we included the following covariates in the model for  the ordinal treatment at visit $j$: visit indicators; status of one or two ARTs and status of HAART  at visits $j-1$ and $j-2$. 
For the denominator in the SIPTW, we additionally included:  square root of CD4 count, $\log_{10}$ of HIV viral load  and HIV symptom scale at visits $j-1$ and $j-2$; the two-way interactions between square root of CD4 count and status of HAART, square root of CD4 count and status of one or two ARTs, $\log_{10}$ of HIV viral load  and status of HAART, $\log_{10}$ of HIV viral load and status of one or two ARTs, square root of CD4 count and $\log_{10}$ of HIV viral load  at visit $j-1$; square root of CD4 count, $\log_{10}$ of HIV viral load and status of one or two ARTs at enrolment; site indicators; and race indicators (black, white, other). 

For IPTCW, a logistic model with the same covariates as those in the treatment assignment model was used for the censoring process to obtain maximum likelihood estimates of the inverse probability of censoring weights and to derive restrictions~\eqref{missingrestrict} for their calibrated version. For fair comparison, the maximum likelihood estimates of the weights in IPTW  were scaled to sum to the sample size available at each visit; and the weights from maximum likelihood for IPTCW were scaled to sum to 5 times the sample size at visit 7 (i.e., the number of outcome measurements that would have been observed had nobody been censored from visit 7 onwards). Finally, we estimated standard errors with 2500 non-parametric bootstrap samples by treating patients as the resampling unit.

\subsection{Results}
We first examine the estimated weights from maximum likelihood and the proposed calibration approach.
Before scaling, the mean of the weights from maximum likelihood for IPTW was 1.01,  and the standard deviation,  the minimum and maximum of these  weights were  0.65, 0.01 and 9.60, respectively. Therefore, these estimated weights do not strongly indicate nonpositivity \citep{Cole2008}, but do suggest that confounding by observed covariates is present in the HERS data. An analysis based on the assumption of no measured confounders, e.g., estimating the parameters of the MSMs with no weighting applied, is thus unlikely to be unbiased. Overall, the empirical distributions of the weights from both maximum likelihood and the calibration approach appear to be well-behaved.  Further details can be found in the Supplementary Material. 

Table~\ref{ResultsTable} presents the  estimates and standard errors of the parameters in the specified MSMs with no weighting, IPTW and IPTCW.
The results of the na\"{i}ve analysis with no weighting applied, as shown in the first two rows of Table~\ref{ResultsTable}, strongly suggest that, compared with no treatment, HAART was effective at increasing the CD4 counts over time for those with $\text{CD4}\leq 500$ at visit 7, and one or two ARTs was effective for those with $200\leq\text{CD4}\leq 500$ at visit 7. However, point estimates for the group with  $\text{CD4}> 500$ at visit 7 showed detrimental effects of both HAART and one or two ARTs. %This could be explained by the presence of strong selection bias among those with $\text{CD4}> 500$ at visit 7, since those with lower CD4 counts are much more likely to receive HAART or ARTs, and previous CD4 counts are highly correlated with current CD4 counts. In contrast, selection bias is less marked in those with $\text{CD4}\leq 500$ as treatment with HAART or ARTs is relatively uniform.  

%The other results can only be regarded as suggestive because they have too much uncertainty. 

\begin{table}[h!]
\centering
\caption{Parameter estimates and their standard errors of the MSMs by applying no weighting,  IPTW and IPTCW with weights from  maximum likelihood (MLE) and from the calibration approach (CMLE) to the HERS data.}
\label{ResultsTable}
\small
\begin{tabular}{cccccc}
\hline
\hline
\multicolumn{1}{c}{Weight}&\multicolumn{1}{c}{Cumulative}&\multicolumn{3}{c}{Strata by CD4 cell count at visit 7}& No stratification\\

\multicolumn{1}{c}{Estimation}&\multicolumn{1}{c}{Effect} &$<200$ & 200-500 &$>500$ & \\
\hline\vspace{0.1in}
&\multicolumn{5}{c}{\textit{No Weighting}} \\
&\multicolumn{1}{r}{$\le 2$  ARTs}&8.57 (9.33)& 13.66 (7.86) & $-$27.36 (18.40) &0.51 (8.06)\\
&\multicolumn{1}{r}{HAART}& 26.34 (8.37)& 27.40 (8.25)& $-$25.59 (16.45)& 14.46 (7.99) \\
&\multicolumn{5}{c}{} \\
&\multicolumn{5}{c}{\textit{Treatment only}} \\
\multicolumn{1}{c}{MLE} && & & &\\
&\multicolumn{1}{r}{$\le 2$ ARTs}&13.27 (9.84) &16.23 (8.71)& $-$26.44 (23.06)&5.59  (9.58) \\
&\multicolumn{1}{r}{HAART}&  27.78 (9.69) & 28.63 (10.24) & $-$2.67 (23.16)&20.89 (10.04)\\
\multicolumn{1}{c}{CMLE}&& & & &\\
&\multicolumn{1}{r}{$\le 2$  ARTs}&14.35 (9.09) &26.60 (7.60)&  5.25 (18.09) & 18.59 (7.23)\\
&\multicolumn{1}{r}{HAART}&  36.53 (8.09) & 34.73 (7.88) & $-$2.75 (17.87)  &  28.16 (7.36)\\
&\multicolumn{5}{c}{} \\
&\multicolumn{5}{c}{\textit{Treatment and dropout}} \\
\multicolumn{1}{c}{MLE} & && & &\\
&\multicolumn{1}{r}{$\le 2$  ARTs} &11.70 (9.29) &  17.11 (8.57) &  $-$24.75 (22.74)& 6.84  (9.07)\\
&\multicolumn{1}{r}{HAART}& 25.79 (9.19)  & 28.80 (10.49)   &  $-$2.08 (22.39)& 21.19 (9.72)\\
\multicolumn{1}{c}{CMLE} && & & &\\
&\multicolumn{1}{r}{$\le 2$  ARTs}&11.92 (8.67) & 27.74 (7.66)& 8.60 (17.74) &19.26 (7.08) \\
&\multicolumn{1}{r}{HAART}& 33.11 (7.93) & 32.75 (8.00) & 3.10 (16.94) & 27.37 (7.24) \\
\hline
\end{tabular}
\end{table}

Applying IPTW with weights from  maximum likelihood provides an upward adjustment of the treatment effects, as seen in the third and fourth rows of Table~\ref{ResultsTable}. The largest adjustments for one or two ARTs and HAART are in the $\text{CD4}<200$ and $\text{CD4}>500$ strata, respectively. Overall, this results in a fairly substantial upward adjustment for the treatment effects in the MSM with no stratification. However, applying IPTW also increased the standard errors of the estimated treatment effects.   

The fifth and sixth rows in Table~\ref{ResultsTable} present the results from applying IPTW with calibrated weights. It appears that  HAART had  an even greater  effect on increasing CD4 counts for those with $\text{CD4}\leq 500$ at visit 7  and overall without stratification, compared with the results based on weights from maximum likelihood. There were also substantial increases in the estimated effects of one or two ARTs for those with $\ge 200$ and overall. As anticipated, the estimated standard errors  with  the calibrated weights are  much smaller even compared to the na\"{i}ve analysis with no weighting applied.

It is possible to gauge how well the weights from maximum likelihood  adjust for confounding from observed covariates, 
by examining the standard deviation of the estimated calibration functions $c(\cdot,\hat{\blambda})$.  A large non-zero value would provide evidence that substantial residual confounding still exists after weighting with weights from maximum likelihood.   For IPTW, the mean and standard deviation of $c(\cdot,\hat{\blambda})$ were  1.09 and 0.61, which suggest that a fair amount of residual confounding from  observed covariates has been addressed after applying the calibration to the weights from maximum likelihood.

Further adjustment for selection bias due to dependent censoring appears to have largely minor effects, as seen in the last four rows of Table~\ref{ResultsTable}. The most notable modifications occur in the $\text{CD4}>500$ strata. However, there is substantial uncertainty associated with these estimated treatment effects, therefore the evidence is insufficient to draw a conclusion. 

As expected, our estimated treatment effects for HAART are generally much larger (more than 1 standard error) than those reported in \cite{Ko2003}, since we have separated the group with one or two ARTs from the group with no treatment. The slightly larger effect of HAART in the $\text{CD4}>500$ strata from  \cite{Ko2003} is again associated with substantial uncertainty.

In conclusion, the results in Table~\ref{ResultsTable} indicate that there were clinically substantial and statistically significant therapeutic effect of cumulative exposure to HAART for those patients with initial CD4 count $\le 500$, which is consistent with the findings in   \cite{Ko2003} and the recommended treatment guideline during the study period of the HERS.  

%For efficiency considerations, the results in Table~\ref{ResultsTable} suggest that it might be (slightly) more beneficial to apply IPTCW instead of IPTW even when censoring is believed to be %non-informative or even when no censoring occurs; the restrictions in~\ref{generalmissingrestrict} are well defined when $C_{ij}=1$ for all $i$ and $j$. A possible explanation is that %calibrated weights for IPTW are designed to estimate causal effects, but do not guarantee for any finite sample that these causal effects are defined for the target population of interest, %thus resulting in conditional (on the covariates) bias. When averaged over different randomization permutations, this becomes variance. In contrast, the calibrated weights for IPTCW %reduces conditional bias by satisfying  additional restrictions that constrain certain weighted sample moments of the joint distribution of covariates to be equivalent to their observed sample %values. For example, \cite{Chan2016} proposed weighted estimators for binary treatments that involve imposing exact three-way balance on the observed covariates among the treated, %control, and the combined group. Chan et al. showed that their estimators are globally efficient. Note, the improved efficiency of IPTCW relative to IPTW was not seen in the simulation %studies because IPTW was applied to larger data sets.

\section{Conclusion and discussion}
In this paper we have proposed a new approach to improving efficiency and robustness of the  IPWE when addressing time-varying confounding and dependent censoring in MSMs for longitudinal outcomes with arbitrary marginal treatment distributions.  Our key idea was to calibrate a set of  initial weights from maximum likelihood by imposing covariate balancing  restrictions that imply treatment assignments are unassociated with histories of covariates and outcomes conditional on treatment history, and the uncensored observations are a representative sample of the target population after weighting the study sample. Our method resembles the use of calibration to improve estimation efficiency in the survey sampling literature, where sampling weights are calibrated to make use of known population information on some auxiliary variables \cite[]{Deville1992}. Specifically, our method calibrates the initial weights to make use of known properties of the true probabilities of treatment assignment and censoring, in particular, their balancing score property \citep{Rosenbaum1983}. Consistent with the empirical and theoretical findings in  the covariate balancing weight literature for the cross-sectional settings, our simulations showed that in longitudinal settings the IPWE  with  calibrated weights had smaller variance and mean squared error than the IPWE with weights from maximum likelihood under correct and incorrect model specification. To the best of our knowledge, our proposed method is the first approach to accommodating both time-varying confounding and dependent censoring in MSMs with arbitrary marginal treatment distributions using the general idea of covariate balancing. As briefly discussed, the difficulty of using  the covariate balancing  weights in longitudinal settings is that it is not obvious how to derive the covariate balancing restrictions in order to improve the estimation of the causal treatment effects of interest. We provided a coherent framework to  derive such restrictions that were tailored to the common scenarios in fitting MSMs using observational cohort data from clinical studies such as the HERS. This will hopefully promote more widespread use of MSMs for various types of treatments/exposure  in practice. 

%We then applied our proposed estimators to the HERS data. This resulted in estimated treatment effects that were further away from the null and with standard errors that were significantly smaller relative to their standard counterparts. To the best of our knowledge, this is the first application where covariate balancing weights for a non-binary treatment have been used to account for confounding and selection bias in a real longitudinal study.     

There are several  directions for future research. First, it would be useful to incorporate data-adaptive methods into our approach. For example, 
 the initial weights from maximum likelihood can be replaced with weights estimated by data-adaptive methods. This can provide some protection from severe model misspecification (e.g., omission of higher-order moments and interactions of the covariates), and therefore reduce the possibility of large bias for IPWEs with calibrated weights. In addition,  data-adaptive methods are also useful to identify functionals of the covariates to be balanced according to whether they  predict the outcome.
Second, it is natural to extend our method to a continuous-time censoring process. This would preclude the need to discretize continuous censoring times. Finally,  our method requires the development of sensitivity analysis strategies to  assess the impact of violations to the no unmeasured confounders assumption. \cite{Ko2003} implemented the sensitivity analysis approach suggested in \cite{Robins1999b} by introducing a sensitivity parameter defined as the difference between the means of the potential outcomes  given observed treatment/covariate histories. This approach is relatively straightforward for binary treatments. But it is not obvious how to generalize it  to non-binary treatments.   
A recent alternative sensitivity analysis approach for IPWEs via percentile bootstrap \cite[]{Zhao2017} may shed some light on this  problem.

\begin{center}
{\bf{\large Supplementary Material}}
\end{center}
Supplementary material includes  further examples of the proposed restrictions, proofs in Sections 4 and 5  as well as additional analyses of the HERS data. 
R code for the simulation study is available at \texttt{https://github.com/seanyiu5/}.

\begin{center}
{\bf{\large Acknowledgements}}
\end{center}
The authors would like to thank Dr Shaun Seaman for helpful comments and discussions. 
Data from the HERS were collected under grant U64-CCU10675 from the U.S. Centers for Disease Control and Prevention. This work is supported by 
 the U.K. Medical Research Council  [Unit programme number MC\_UU\_00002/8; grant number MR/M025152/1].

\baselineskip=16pt
\bibliographystyle{rss}

\end{document}

\maketitle

\baselineskip=24pt
\setlength{\parindent}{.25in}

\maketitle
\section{Calibration restrictions for  time-varying continuous treatments}
We demonstrate how restrictions in equation (4) of the main text can be derived for a sequence of continuous treatments. We assume that the time-varying continuous treatment at visit $j$ for the $i$th patient follows a  heteroscedastic normal linear model  $A_{ij}\sim N\{\widetilde{\bX}_{i,j-1}^{\mu\top}\bbeta^{\mu},\exp(\widetilde{\bX}_{i,j-1}^{\sigma\top}\bbeta^{\sigma})\}$, where $\widetilde{\bX}_{i,j-1}^{\mu}$ and $\widetilde{\bX}_{i,j-1}^{\sigma}$ include $1$ and functionals of  $\overline{X}_{i,j-1}$ (e.g., interactions), and 
$\bbeta^{\mu}$ and $\bbeta^{\sigma}$ are corresponding regression coefficients. Then
\begin{eqnarray}
\frac{\partial}{\partial \bbeta^{\mu}} \log\{\text{pr}(A_{ij}\mid \widetilde{\bX}_{i,j-1}^{\mu}, \widetilde{\bX}_{i,j-1}^{\sigma};\bbeta^{\mu}, \bbeta^{\sigma})\}&=&\frac{(A_{ij}-\widetilde{\bX}_{i,j-1}^{\mu\top}\bbeta^{\mu})}{\exp(\widetilde{\bX}_{i,j-1}^{\sigma\top}\bbeta^{\sigma})}\widetilde{\bX}_{i,j-1}^{\mu}, \nn\\
\frac{\partial}{\partial \bbeta^{\sigma}} \log\{\text{pr}(A_{ij}\mid \widetilde{\bX}_{i,j-1}^{\mu}, \widetilde{\bX}_{i,j-1}^{\sigma};\bbeta^{\mu}, \bbeta^{\sigma})\}&=&\frac{1}{2}\left\{-1+\frac{(A_{ij}-\widetilde{\bX}_{i,j-1}^{\mu\top}\bbeta^{\mu})^2}{\exp(\widetilde{\bX}_{i,j-1}^{\sigma\top}\bbeta^{\sigma})}\right\}\widetilde{\bX}_{i,j-1}^{\sigma}. \nn
\end{eqnarray}

By substituting the above equations into (4) of the main text, the restrictions based on this normal linear model for the continuous  treatments are
\begin{equation*}
\begin{split}
\sum_{j=1}^T \sum_{i=1}^n SW^{A\star}_{ij}({\blambda})\sum_{k=1}^j\frac{(A_{ik}-\hat{\mu}_{ik})}{\hat{\sigma}^2_{ik}}\widetilde{\bX}_{i,k-1}^{\mu}=0,\\
\sum_{j=1}^T\sum_{i=1}^n SW^{A\star}_{ij}({\blambda})\sum_{k=1}^j\left\{-1+\frac{(A_{ik}-\hat{\mu}_{ik})^2}{\hat{\sigma}^2_{ik}}\right\}\widetilde{\bX}_{i,k-1}^{\sigma}=0,
\end{split}
\end{equation*}
where $\hat{\mu}_{ik}$ and $\hat{\sigma}^2_{ik}$ are the estimated mean and variance of the continuous treatments from fitting the same normal linear model but with treatment history as the only covariates. Similarly to the ordinal treatments in the main text, these restrictions are designed to remove the associations between the covariates ($\widetilde{\bX}_{i,k-1}^{\mu}$ and $\widetilde{\bX}_{i,k-1}^{\sigma}$)  and the standardized residuals from the treatment model for the numerator of the initial stabilized weights after weighting the current sample. 

%where $\hat{\mu}_{ik}=(\hat{\alpha}_{\mu},\hat{\alpha}_{\sigma})$ are obtained by fitting the model $A_{ik}\sim N\{\alpha^\top_{\mu}[1,\overline{A}_{ik-1}],\exp(\alpha^\top_{\sigma}[1,\overline{A}_{ik-1}])\}$ to the observed treatment assignment data.

\section{Property of inverse probability of censoring weights}
We show that $\text{pr}^*(R_{i1}\mid  \overline{H}_{i0})=1/2$ in Section 4.2.1 of the main text, where $*$ denotes the pseudo-population consisting of  the $1/\text{pr}(R_{i1}=1| \overline{H}_{i0})-1$  copies of  the complete cases (i.e., those with $R_{i1}=1$) and the incomplete cases (i.e., those with $R_{i1}=0$).
\begin{proof}
Specifically, we have
\begin{equation*}
\begin{split}
\text{pr}^*(R_{i1}| \overline{H}_{i0})&=\frac{R_{i1}\{1/\text{pr}(R_{i1}=1| \overline{H}_{i0})-1\}\text{pr}(R_{i1}=1| \overline{H}_{i0})+(1-R_{i1})\text{pr}(R_{i1}=0| \overline{H}_{i0})}{\{1/\text{pr}(R_{i1}=1| \overline{H}_{i0})-1\}\text{pr}(R_{i1}=1| \overline{H}_{i0})+\text{pr}(R_{i1}=0| \overline{H}_{i0})}\\
&=\frac{R_{i1}\text{pr}(R_{i1}=0| \overline{H}_{i0})+(1-R_{i1})\text{pr}(R_{i1}=0| \overline{H}_{i0})}{2\text{pr}(R_{i1}=0| \overline{H}_{i0})}\\
&=1/2
\end{split}
\end{equation*} 
\end{proof}

\section{Equivalence of censoring restrictions in (16) and (17) of the main text}\label{Identity_for_restrictions}
By writing out the terms in the summation in $j$, equation (16) of the main text is equal to
\begin{align*}
&\sum_{i=1}^n\Big[T\{R_{i1}W^{C\star}_{i1}(
\blambda)-1\}\widetilde{\bH}_{i0}+(T-1)\{R_{i2}W^{C\star}_{i2}(\blambda)-R_{i1}W^{C\star}_{i1}(\blambda)\}\widetilde{\bH}_{i1}\\
+&\ldots +\{R_{iT}W^{C\star}_{iT}(\blambda)-R_{i,T-1}W^{C\star}_{i,T-1}(\blambda)\}\widetilde{\bH}_{i,T-1}\Big]=0,
\end{align*}
which follows from the fact that $R_{i0}=1$ for all $i$ and by convention $W^{C\star}_{i0}(\blambda)=1$ for all $i$. Next, we bring $T\sum_{i=1}^n\widetilde{\bH}_{i0}$ to the right-hand side,
\begin{align*}
&\sum_{i=1}^n\Big[TR_{i1}W^{C\star}_{i1}(
\blambda)\widetilde{\bH}_{i0}+(T-1)\{R_{i2}W^{C\star}_{i2}(\blambda)-R_{i1}W^{C\star}_{i1}(\blambda)\}\widetilde{\bH}_{i1}\\
+&\ldots +\{R_{iT}W^{C\star}_{iT}(\blambda)-R_{i,T-1}W^{C\star}_{i,T-1}(\blambda)\}\widetilde{\bH}_{i,T-1}\Big]=T\sum_{i=1}^n\widetilde{\bH}_{i0}.
\end{align*}
Now collect the coefficients for the weights
\begin{align*}
&\sum_{i=1}^n\Big[R_{i1}W^{C\star}_{i1}(\blambda)\{T\widetilde{\bH}_{i0}-(T-1)\widetilde{\bH}_{i1}\}+R_{i2}W^{C\star}_{i2}(\blambda)\{(T-1)\widetilde{\bH}_{i1}-(T-2)\widetilde{\bH}_{i2}\}\\
+&\ldots +R_{iT}W^{C\star}_{iT}(\blambda)\widetilde{\bH}_{i,T-1}\Big]=T\sum_{i=1}^n\widetilde{\bH}_{i0}.
\end{align*}
When written in summation form, the above equation is equivalent to (17) in the main text.

\section{Calibration restrictions for stabilized inverse probability of censoring weights}
Similarly to inverse probability of treatment weighting,  stabilized weights for censoring, $SW^C_{ij}(\hat{\bet},\hat{\btheta})=W^C_{ij}(\hat{\btheta})\prod_{k=1}^j\pi^{\text{s}}_{ik}(\hat{\bet})$, can be used, where $\pi^{\text{s}}_{ik}(\bet)=\text{pr}(R_{ik}=1\mid \overline{A}_{i,k-1},R_{i,k-1}=1;{\bet})$ is a parametric model for the censoring process given the treatment history only. For completeness, we describe our proposed method for calibrating stabilized weights for censoring in this section. 

\subsection{The two time period setting}
The idea of stabilized weights is to create a pseudo-population that is representative of the population that would have been observed had nobody been censored at visit 1, which we refer to as the complete population,  and where each patient in the complete population has been weighted by $\pi^{\text{s}}_{i1}(\hat{\bet})$. 
In other words,  we down-weight patients who are unlikely to be observed at visit 1 based solely on treatment history. Therefore the task is less ambitious than what unstabilized weights are trying to achieve. Restrictions for the stabilized weights  can be derived, e.g.,  by multiplying the summands in (10) of the main text by $\pi^{\text{s}}_{i1}(\hat{\bet})$, 
$$ \sum_{i=1}^n R_{i1}SW^{C\star}_{i1}({\blambda})\widetilde{\bH}_{i0}=\sum_{i=1}^n\pi^{\text{s}}_{i1}(\hat{\bet})\widetilde{\bH}_{i0}, $$ where
$SW^{C\star}_{i1}({\blambda})=SW^C_{i1}(\hat{\bet},\hat{\btheta})c(\overline{H}_{i0},\blambda)$. Note that, if the censoring model is correctly specified, i.e., $W^C_{i1}(\hat{\btheta})$ converges in probability to the true censoring weight $W^{C}_{i1}$ as $n\rightarrow \infty$, and $\Ex(R_{i1}W^{C}_{i1}\mid \overline{H}_{i0},R_{i0}=1)=1$, then these restrictions, after scaling by $n$, are satisfied asymptotically.

\subsection{The general longitudinal setting}
Restrictions for stabilized censoring weights in the general longitudinal setting can be derived similarly as above. In particular, we can multiply the summand indexed by $j$ in equation (16) of the main text by the stabilizing factor $\prod_{k=1}^j\pi^{\text{s}}_{ik}(\hat{\bet})$, 
\begin{equation*}
\sum_{j=1}^T \sum_{i=1}^n (T-j+1)\left[R_{ij}SW^{C\star}_{ij}({\blambda})-R_{i,j-1}SW^{C\star}_{i,j-1}({\blambda})\pi^{\text{s}}_{ij}(\hat{\bet})\right]\widetilde{\bH}_{i,j-1}=0,
\end{equation*}
where $SW^{C\star}_{ij}({\blambda})=SW^C_{ij}(\hat{\bet},\hat{\btheta})c(\overline{H}_{i,j-1},\blambda)$. These restrictions, after scaling by $n$, are again satisfied asymptotically $(n\rightarrow \infty)$ without calibration if the censoring model is  correctly specified, i.e., when $W^C_{ij}(\hat{\btheta})$ converges in probability to the true censoring weight $W^{C}_{ij}$ as $n\rightarrow \infty$, and $\Ex(R_{ij}W^{C}_{ij}\mid \overline{H}_{i,j-1}, R_{i0}=1)=1$. %\tp{I am not entirely follow the last equation. $W^{C}_{ij}$ is a cumulative weight up to visit j. Should there be a product of $R_{ij}$ here?}

To facilitate the implementation procedure, it is more convenient to re-express these restrictions as
\begin{equation*}
\sum_{j=1}^T  \sum_{i=1}^nR_{ij}SW^{C\star}_{ij}({\blambda})\left[(T-j+1)\widetilde{\bH}_{i, j-1}-(T-j)\pi^{\text{s}}_{i,j+1}(\hat{\bet})\widetilde{\bH}_{ij}\right]=T\sum_{i=1}^n\pi^{\text{s}}_{i1}(\hat{\bet})\widetilde{\bH}_{i0},
\end{equation*} 
which can be derived by using the technique described in Section~\ref{Identity_for_restrictions}.   
%\tp{ When $j=T$,$\pi^{\text{s}}_{i,j+1}$ is not defined, is it?}
 
\section{The impact of calibration on the true inverse probability of treatment weights}
We show that the true  inverse probability of treatment weights satisfy the restrictions (4) in the main text asymptotically, so calibration will not alter these weights asymptotically. As a result,  calibration will maintain the consistency of the inverse probability of treatment weighted estimator when the models for the initial weights are correctly specified. 

Without loss of generality, suppose that $\overline{A}_{ij}$ and $\overline{X}_{i,j-1}$ $(j=1,\ldots,T)$ are continuous. Let 
$SW^{A}_{ij}(\balpha)=\prod_{k=1}^j\text{pr}(A_{ik}\mid \overline{A}_{i,k-1};\balpha)/\prod_{k=1}^j\text{pr}(A_{ik}\mid \overline{X}_{i,k-1})$. That is,  the numerator in $SW^{A}_{ij}(\balpha)$ is arbitrary and only depends on treatment history,  and the denominator in $SW^{A}_{ij}(\balpha)$ is  the true treatment assignment probabilities. 
We show that 
\begin{equation*}
\Ex\left[SW^{A}_{ij}(\balpha)\sum_{k=1}^j\frac{\partial}{\partial \bbeta_\text{w}} \log\{\text{pr}(A_{ik}\mid \overline{X}_{i,k-1};\bbeta_\text{w})\}\Big|_{\{{\footnotesize \bbeta}_\text{wb}={\footnotesize \balpha},{\footnotesize \bbeta}_\text{wd}=\mathbf{0}\}}\right]=0
\end{equation*}
for any model $\text{pr}(A_{ik}\mid \overline{X}_{i,k-1};\bbeta_\text{w})$ that satisfies $\text{pr}(A_{ik}\mid \overline{X}_{i,k-1};\bbeta_{\text{wb}}=\balpha,\bbeta_{\text{wd}}=\mathbf{0})=\text{pr}(A_{ik}\mid \overline{A}_{i,k-1};\balpha)$.

\begin{proof}
\begin{equation*}
\begin{split}
&\Ex\left[SW^{A}_{ij}(\balpha)\sum_{k=1}^j\frac{\partial}{\partial \bbeta_\text{w}} \log\{\text{pr}(A_{ik}\mid \overline{X}_{i,k-1};\bbeta_\text{w})\}\Big|_{\{{\footnotesize \bbeta}_\text{wb}={\footnotesize \balpha},{\footnotesize \bbeta}_\text{wd}=\mathbf{0}\}}\right]\\
=&\Ex\left[\prod_{k=1}^j\frac{\text{pr}(A_{ik}\mid \overline{A}_{i,k-1};\balpha)}{\text{pr}(A_{ik}\mid \overline{X}_{i,k-1})}\left\{\frac{\partial}{\partial \bbeta_\text{w}} \log\left\{\prod_{k=1}^j\text{pr}(A_{ik}\mid \overline{X}_{i,k-1};\bbeta_\text{w})\right\}\Big|_{\{{\footnotesize \bbeta}_\text{wb}={\footnotesize \balpha},{\footnotesize \bbeta}_\text{wd}=\mathbf{0}\}}\right\}\right]\\
=&\Ex\left[\prod_{k=1}^j\frac{1}{\text{pr}(A_{ik}\mid \overline{X}_{i,k-1})}\left\{\frac{\partial}{\partial \bbeta_\text{w}} \prod_{k=1}^j\text{pr}(A_{ik}\mid \overline{X}_{i,k-1};\bbeta_\text{w})\Big|_{\{{\footnotesize \bbeta}_\text{wb}={\footnotesize \balpha},{\footnotesize \bbeta}_\text{wd}=\mathbf{0}\}}\right\}\right]\\
=&\int_{\overline{A}_{ij}}\int_{\overline{X}_{i,j-1}}\prod_{k=1}^j\frac{1}{\text{pr}(A_{ik}\mid \overline{X}_{i,k-1})}\left\{\frac{\partial}{\partial \bbeta_\text{w}} \prod_{k=1}^j\text{pr}(A_{ik}\mid \overline{X}_{i,k-1};\bbeta_\text{w})\Big|_{\{{\footnotesize \bbeta}_\text{wb}={\footnotesize \balpha},{\footnotesize \bbeta}_\text{wd}=\mathbf{0}\}}\right\}\\
\times&\prod_{k=1}^j\left\{\text{pr}(A_{ik}\mid\overline{X}_{i,k-1})\text{pr}(X_{i,k-1}\mid \overline{A}_{i,k-1},\overline{X}_{i,k-2})\right\}\,d\overline{A}_{ij}\,d\overline{X}_{i,j-1}\\
=&\int_{\overline{A}_{ij}}\int_{\overline{X}_{i,j-1}}\left\{\frac{\partial}{\partial \bbeta_\text{w}} \prod_{k=1}^j\text{pr}(A_{ik}\mid \overline{X}_{i,k-1};\bbeta_\text{w})\Big|_{\{{\footnotesize \bbeta}_\text{wb}={\footnotesize \balpha},{\footnotesize \bbeta}_\text{wd}=\mathbf{0}\}}\right\}\\
\times&\prod_{k=1}^j\text{pr}(X_{i,k-1}\mid \overline{A}_{i,k-1},\overline{X}_{i,k-2})\,d\overline{A}_{ij}\,d\overline{X}_{i,j-1}.\\
\end{split}
\end{equation*}
Assuming that we can interchange differentiation and integration, the above expression is equal to
\begin{equation*}
\begin{split}
&\frac{\partial}{\partial \bbeta_\text{w}}\left\{\int_{\overline{A}_{ij}}\int_{\overline{X}_{i,j-1}} \prod_{k=1}^j\text{pr}(A_{ik}\mid \overline{X}_{i,k-1};\bbeta_\text{w})\text{pr}(X_{i,k-1}\mid \overline{A}_{i,k-1},\overline{X}_{i,k-2})\,d\overline{A}_{ij}\,d\overline{X}_{i,j-1}\right\}\Big|_{\{{\footnotesize \bbeta}_\text{wb}={\footnotesize \balpha},{\footnotesize \bbeta}_\text{wd}=\mathbf{0}\}}\\
=&\frac{\partial}{\partial \bbeta_\text{w}}1\Big|_{\{{\footnotesize \bbeta}_\text{wb}={\footnotesize \balpha},{\footnotesize \bbeta}_\text{wd}=\mathbf{0}\}}=0.
\end{split}
\end{equation*}

\end{proof}

\section{Asymptotic covariate imbalances for the maximum likelihood approach under model misspecification}\label{imbalancesection}
In this section, we briefly discuss why the maximum likelihood approach may perform poorly in removing asymptotic imbalances of covariate distributions even under mild model misspecification. For clearer exposition, we focus on the cross-sectional setting, i.e., we drop the visit subscript. Following   the weighting framework in \cite{Yiu2017}, we can use   the following measure to assess the  covariate balance in the pseudo-population after weighting,
\[\frac{1}{n}\sum_{i=1}^n SW^{A}_i(\hat{\balpha},\hat{\bbeta})\frac{\partial}{\partial \bbeta_\text{w}} \log\{\text{pr}(A_i\mid X_i;\bbeta_\text{w})\}\Big|_{{\{{\footnotesize \bbeta}_{\text{wb}}=\hat{\footnotesize \balpha},{\footnotesize \bbeta}_{\text{wd}}=\mathbf{0}\}}},\]
which is equal to the zero vector if and only if covariates are balanced in the current sample. Without loss of generality, let $A_i$ and $X_i$ be continuous, and $\balpha^{*}$ and $\bbeta^{*}$ be the probability limits of $\hat{\balpha}$ and $\hat{\bbeta}$, then this measure is asymptotically equivalent to
\begin{equation}\label{imbalance}
\begin{split}
&\Ex\left(SW^{A}_i(\balpha^{*},\bbeta^{*})\frac{\partial}{\partial \bbeta_\text{w}} \log\{\text{pr}(A_i\mid X_i;\bbeta_\text{w})\}\Big|_{{\{{\footnotesize \bbeta}_{\text{wb}}={\footnotesize\balpha}^{*},{\footnotesize \bbeta}_{\text{wd}}=\mathbf{0}\}}}\right)\\
=&\int_{X_i}\int_{A_i}\frac{\text{pr}(A_i;\balpha^{*})}{\text{pr}(A_i\mid X_i;\bbeta^{*})}\frac{\partial}{\partial \bbeta_\text{w}} \log\{\text{pr}(A_i\mid X_i;\bbeta_\text{w})\}\Big|_{{\{{\footnotesize \bbeta}_{\text{wb}}={\footnotesize\balpha}^{*},{\footnotesize \bbeta}_{\text{wd}}=\mathbf{0}\}}}~\text{pr}(A_i, X_i)~dA_i~dX_i\\
=&\int_{X_i}\int_{A_i}\frac{\text{pr}(A_i \mid X_i)}{\text{pr}(A_i\mid X_i;\bbeta^{*})}\frac{\partial}{\partial \bbeta_\text{w}} \text{pr}(A_i\mid X_i;\bbeta_\text{w})\Big|_{{\{{\footnotesize \bbeta}_{\text{wb}}={\footnotesize\balpha}^{*},{\footnotesize \bbeta}_{\text{wd}}=\mathbf{0}\}}}~\text{pr}(X_i)~dA_i~dX_i\\
=&\int_{X_i}\int_{A_i}\left(\frac{\text{pr}(A_i \mid X_i)}{\text{pr}(A_i\mid X_i;\bbeta^{*})}-1\right)\frac{\partial}{\partial \bbeta_\text{w}} \text{pr}(A_i\mid X_i;\bbeta_\text{w})\Big|_{{\{{\footnotesize \bbeta}_{\text{wb}}={\footnotesize\balpha}^{*},{\footnotesize \bbeta}_{\text{wd}}=\mathbf{0}\}}}~\text{pr}(X_i)~dA_i~dX_i,
\end{split}
\end{equation}
where the last line follows from the fact that under standard regularity assumptions, i.e., we can interchange differentiation and integration, 
\begin{equation*}
\int_{X_i}\int_{A_i}\frac{\partial}{\partial \bbeta_\text{w}} \text{pr}(A_i\mid X_i;\bbeta_\text{w})\Big|_{{\{{\footnotesize \bbeta}_{\text{wb}}={\footnotesize\balpha}^{*},{\footnotesize \bbeta}_{\text{wd}}=\mathbf{0}\}}}~\text{pr}(X_i)~dA_i~dX_i=0.
\end{equation*}
From \eqref{imbalance}, it is clear that small asymptotic imbalances can be attained whenever the relative errors of the treatment probabilities are small, i.e., $|\text{pr}(A_i \mid X_i)/\text{pr}(A_i\mid X_i;\bbeta^{*})-1|\approx 0$. Methods for estimating $\bbeta$ that are targeted to achieving small absolute errors of the treatment probabilities, i.e., $|\text{pr}(A_i \mid X_i)-\text{pr}(A_i\mid X_i;\bbeta^{*})|\approx 0$, under mild model misspecification, such as least squares, are therefore not guaranteed to perform well in terms of reducing covariate imbalances.  This is because, as mentioned in the introduction of the main text, small absolute errors do not necessarily imply small relative errors. Unfortunately, maximum likelihood for weight estimation is more geared towards achieving small absolute errors through minimizing the Kullback-Leibler divergence rather than small relative errors \cite[]{Tan2017}.
As a result, the asymptotic  covariate imbalances possibly from weights by maximum likelihood can  lead to poor performance of the corresponding IPWE.

\section{Additional results from the HERS analysis}
\begin{figure}
\includegraphics{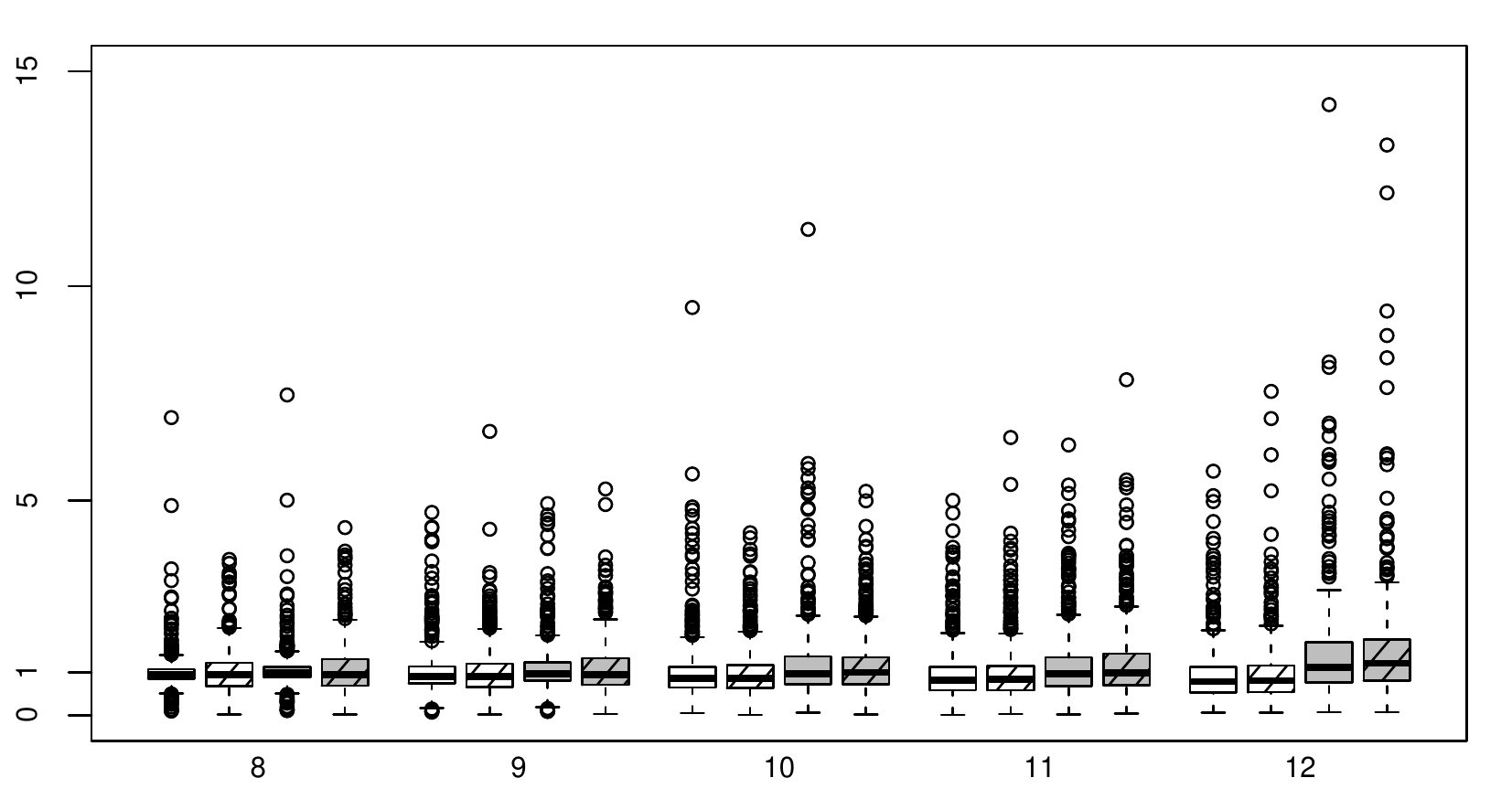}
\centering
\caption{Box plots of the estimated weights  stratified by visit  in the HERS analysis in Section 7 of the main text.  Left two white boxes: for treatment only;  right two gray boxes:   for treatment and dropout;  unshaded boxes: weights  based on maximum likelihood; shaded boxes: calibrated weights.} 
\label{Fig1}
\end{figure}

Figure \ref{Fig1} displays boxplots of the estimated weights by visit in the HERS analysis. In all cases, the weights appear to be fairly well behaved and no weight seems to exert undue influence relative to the other weights, especially for  weighting with treatment only, where  all weights are less than ten.
%The weights are more variable at the later visits.

In addition to the marginal structural model  specified in the main text, we also fit  the  following marginal structural model
\begin{equation}\label{MSM}
\Ex(Y^{\overline{a}_{j}}_{ij})=\delta_{0j}+\sum_{k=1}^2\delta_kI (D_i=k)+\bdelta^\top_{v}\bV_{i}+\sum_{k=0}^2I(D_i=k)\left\{\gamma_{1k}(a^0_{j}-a^1_{j})+\gamma_{2k}a^1_{j}\right\}
\end{equation}
for $j=8,\ldots,12$, where $\delta_{0j}$ are visit-specific intercept terms, $\bV_{i}$ are baseline covariates  evaluated at visit 7, and $\bdelta_v$ are their corresponding regression coefficients, $\gamma_{1k}$ and $\gamma_{2k}$ represent strata-specific causal effects of recent exposures to one or two ARTs and HAART, respectively. This model quantifies the causal effect of receiving one or two ARTs and HAART in the previous six months on the current CD4 count, and therefore reflects the short-term treatment effects.  Again, we also fit a marginal structural model without stratification by CD4 count at visit 7 in \eqref{MSM} . %However, as model \eqref{MSM} does not contain treatment exposures prior to six months, it can not inform on clinically meaningful quantities like the mean CD4 count under one year of HAART exposure relative to no HAART exposure. 

Table \ref{ResultsTable} presents the results from fitting model \eqref{MSM} with  weights estimated by the approaches  described in the main text. Overall, these results exhibit a similar pattern to those reported in Table 4 of the main text, especially for the model without  stratification. Applying inverse probability of treatment weighting with weights from  maximum likelihood  generally provides an upward adjustment to the  estimated treatment effects when  no weighting is applied, although with the cost of losing efficiency. In contrast, the calibration approach provides an even larger upward adjustment relative to the no weighting approach and is also more efficient. The consequences of adjusting for dependent censoring are again largely minor. Finally, the estimated effects of HAART are generally larger than those in \cite{Ko2003}, except for the group CD4 counts $>500$ at visit 7 where there is much uncertainty. 

While  \cite{Ko2003} found therapeutic effects of HAART in this group with CD4 counts $>500$, our analysis suggests that this could be a result of mixing the `no HAART' group and the `one or two ARTs' group, since we found that the effect of HAART only appears to be therapeutic relative to one or two ARTs but not relative to no treatment when applying the maximum likelihood approach. In contrast, the calibration approach suggests that HAART is not even therapeutic relative to one or two ARTs in the group with  CD4 counts $>500$. However, since treatment history is not adjusted for  in the MSM, there could be unmeasured confounding reflected in the treatment history that leads to these counter-intuitive results. 
Moreover, there is  substantial uncertainty associated with the point estimates. 
Therefore, the evidence is not sufficient to draw a conclusion about the treatment effects for the baseline group with CD4 counts $>500$.
% or by allowing the numerator of the stabilized weights to depend on treatment history,
%se results are most sensibly explained by the manifestation of unexplained selection bias in which treatment history can alleviate.  will also likely result in overestimates of the true direct effects of treatment exposure in the previous six months. 

\begin{table}[h!]
\centering
\caption{Parameter estimates and their standard errors of the MSMs in \eqref{MSM} by applying no weighting,  inverse probability of treatment weighting  and inverse probability of treatment and censoring weighting  with weights from  maximum likelihood (MLE) and from the calibration approach (CMLE) to the HERS data.}
\label{ResultsTable}
\small
\begin{tabular}{cccccc}
\hline
\hline
\multicolumn{1}{c}{Weight}&\multicolumn{1}{c}{Cumulative}&\multicolumn{3}{c}{Strata by CD4 cell count at visit 7}& No stratification\\

\multicolumn{1}{c}{Estimation}&\multicolumn{1}{c}{Effect} &$<200$ & 200-500 &$>500$ & \\
\hline\vspace{0.1in}
&\multicolumn{5}{c}{\textit{No Weighting}} \\
&\multicolumn{1}{r}{$\le 2$  ARTs}&23.67 (24.39)&45.60 (18.22)  &-96.08 (43.26)  &-0.88 (16.78)\\
&\multicolumn{1}{r}{HAART}& 67.37 (24.51)&75.86 (19.94)  &-94.15 (46.32)  & 28.93 (18.27)\\
&\multicolumn{5}{c}{} \\
&\multicolumn{5}{c}{\textit{Treatment only}} \\
\multicolumn{1}{c}{MLE} && & & &\\
&\multicolumn{1}{r}{$\le 2$ ARTs}&42.50 (24.86)&48.40 (20.89)  & -62.00 (51.89)  &18.42 (20.10) \\
&\multicolumn{1}{r}{HAART}& 80.92 (25.98)&77.53 (25.21)  &-33.96 (59.05)  &48.85 (23.24)\\
\multicolumn{1}{c}{CMLE}&& & & &\\
&\multicolumn{1}{r}{$\le 2$  ARTs}&25.92 (22.66) & 69.53 (20.36) &-22.36 (42.27)  &35.91 (16.37)\\
&\multicolumn{1}{r}{HAART}&  81.64 (21.79)&87.17 (19.65)  &-38.40 (46.75)  &59.07 (16.32)\\
&\multicolumn{5}{c}{} \\
&\multicolumn{5}{c}{\textit{Treatment and dropout}} \\
\multicolumn{1}{c}{MLE} & && & &\\
&\multicolumn{1}{r}{$\le 2$  ARTs} &38.67 (25.41)&47.90 (21.29)  &-67.29 (51.72)  &17.36 (19.92)\\
&\multicolumn{1}{r}{HAART}& 79.47 (26.56)&77.27 (26.35)  &-36.25 (58.11)  &49.25 (23.27)\\
\multicolumn{1}{c}{CMLE} && & & &\\
&\multicolumn{1}{r}{$\le 2$  ARTs}&25.39 (23.13)&71.23 (21.29)  &-22.00 (42.40)  &35.64 (16.82) \\
&\multicolumn{1}{r}{HAART}& 82.15 (22.75)&83.95 (20.72)  &-27.34 (44.32)  &58.71 (16.85) \\
\hline
\end{tabular}
\end{table}
\newpage
\bibliographystyle{rss}